\documentclass[12pt]{article}
\pdfoutput=1

\usepackage{amsmath}
\usepackage{amssymb}
\usepackage{amsfonts}
\usepackage{graphicx}
\usepackage[utf8]{inputenc}		
\usepackage{aas_macros}				

\usepackage{slashed}				
\usepackage{bbm}					
\usepackage{bm}        

\usepackage[dvipsnames]{xcolor}
\usepackage[hang,flushmargin]{footmisc} 

\usepackage{fancyhdr}		
\usepackage{lipsum}			
\usepackage{framed}			
\usepackage{cite}			

\usepackage{mathtools}

\usepackage{booktabs}		
\usepackage{multirow}		

\usepackage[font=small]{caption} 
\usepackage{float}         

\usepackage[margin=2cm]{geometry}   
\graphicspath{{figures/}}			
\numberwithin{equation}{section}    

\usepackage{multicol}
\usepackage{etoolbox}
\usepackage{relsize}
\patchcmd{\thebibliography}
  {\list}
  {\begin{multicols}{2}\smaller\list}
  {}
  {}
\appto{\endthebibliography}{\end{multicols}}

\let\oldenumerate\enumerate
\renewcommand{\enumerate}{
  \oldenumerate
  \setlength{\itemsep}{1pt}
  \setlength{\parskip}{0pt}
  \setlength{\parsep}{0pt}
}

\let\olditemize\itemize
\renewcommand{\itemize}{
  \olditemize
  \setlength{\itemsep}{1pt}
  \setlength{\parskip}{0pt}
  \setlength{\parsep}{0pt}
}

\newcommand{\acro}[1]{#1}

\renewcommand{\tilde}{\widetilde}   
\renewcommand{\vec}[1]{\bm{#1}} 
\newcommand{\ket}[1]{\left|#1\right\rangle}    
\newcommand{\bra}[1]{\left\langle#1\right|}    

\newcommand{\email}[1]{\href{mailto:#1}{#1}}
\newenvironment{institutions}[1][2em]{\begin{list}{}{\setlength\leftmargin{#1}\setlength\rightmargin{#1}}\item[]}{\end{list}}

\newcommand{\be}{\begin{equation}}
\newcommand{\ee}{\end{equation}}
\newcommand{\bea}{\begin{eqnarray}}
\newcommand{\eea}{\end{eqnarray}}

\usepackage[
	colorlinks=true,
	citecolor=green!50!black,
	linkcolor=NavyBlue!75!black,
	urlcolor=green!50!black,
	hypertexnames=false]{hyperref}

\fancypagestyle{firststyle}{
	\rhead{\footnotesize%
	\texttt{UCR-TR-2019-FLIP-NCC-1709}%
	}}

\usepackage{etoc}

\begin{document}

\thispagestyle{firststyle} 	

\begin{center}

{\Large \bf Exotic Spin-Dependent Forces from a Hidden Sector }

    \vskip .7cm

   { \bf
   	Alexandria~Costantino$^{a}$,
    Sylvain~Fichet$^{\,b,c}$
   	and
   	Philip Tanedo$^{a}$
   	}
   \\
   \vspace{-.2em}
   { \tt \footnotesize
      \email{acost007@ucr.edu},
      \email{sfichet@caltech.edu},
	    \email{flip.tanedo@ucr.edu}
   }

   \vspace{-.2cm}

   \begin{institutions}[.5cm]
   \footnotesize
   $^{a}$
   {\it
	    Department of Physics \& Astronomy,
	    University of California, 900 University Ave, Riverside,
	    USA
	    }
	\\
	\vspace*{0.05cm}
	$^{b}$
	{\it
       Walter Burke Institute for Theoretical Physics,
       California Institute of Technology,
       1200 E California Blvd,
       Pasadena, USA
       }
  \\
 	\vspace*{0.05cm}
 	$^{c}$
 	{\it
        \acro{ICTP} South American Institute for Fundamental Research  \& \acro{IFT-UNESP},
        R.~Dr.~Bento Teobaldo Ferraz 271, S\~ao Paulo, Brazil
        }
   \end{institutions}

\end{center}


\begin{abstract}
\noindent

New dynamics from hidden sectors may manifest as long-range forces between visible matter particles. The well-known case of Yukawa-like potentials occurs via the exchange of a single virtual particle. However, more exotic behavior is also possible. We present three classes of exotic potentials that are generated by relativistic theories: (i) quantum forces from the loop-level exchange of two virtual particles, (ii) conformal forces from a conformal  sector, and (iii)  emergent forces from degrees of freedom that only exist in the infrared regime of the theory. 
We discuss the complementarity of spin-dependent force searches in an effective field theory framework. We identify  well-motivated directions to search for exotic spin-dependent forces. 

\end{abstract}

\small
\setcounter{tocdepth}{1}
\tableofcontents
\normalsize
\clearpage


\section{Introduction}

The existence of dark matter and dark energy suggests the possible existence of a light hidden sector. To avoid experimental observation, the particles in this hidden sector should have suppressed interactions with visible matter; these sectors are broadly referred to as \textit{dark sectors}.
The existence of a dark sector may imply that nature exhibits new macroscopic forces between  visible sector particles. For example the exchange of a single bosonic particle induces a Yukawa-like potential.
A multitude of experimental searches probe the possible existence of spin-independent forces, see {e.g.}~\cite{Adelberger:2003zx,Salumbides:2013dua, Brax:2017xho}.
However, it is also possible that the dominant effects of a hidden sector force could be \textit{spin-dependent}.
These types of forces are more challenging to observe and relatively
few experiments are designed to probe them.

Both theoretical and experimental efforts have focused primarily on Yukawa-like spin-dependent forces that arise from the exchange of a single massive boson. Spin-dependent forces from the axion were identified in~\cite{Moody:1984ba}. 
More recently, Dobrescu and Mociou presented a dictionary between the field theoretical properties of new bosons and the types of spin-dependent macroscopic forces that they generate~\cite{Dobrescu:2006au}. See \cite{Fadeev:2018rfl} for a recent discussion that includes contact interactions, further phenomenology, and corrections of earlier literature.
Conversely, experiments have been focused on the search for Yukawa-like forces, see {e.g.}~\cite{Piegsa:2012th,  Yan:2012wk, Ledbetter:2012xd}.

In this manuscript we present 
\textit{exotic} forces, with a primary focus on spin-dependent ones. 
We define exotic to mean forces that are not Yukawa-like.
The main goal of this paper is to fill a gap in the literature by presenting exotic potentials generated by explicit
dark  sector  models.\,\footnote{See also \cite{Ferrer:1998ue,Fichet:2017bng,  Brax:2017xho} for related work on spin-independent potentials.}
As a preliminary study, we systematically consider the experimental complementarity of exotic and Yukawa-like potentials from an effective field theory perspective.

\begin{figure}
  \centering
  \includegraphics[width=0.9\textwidth]{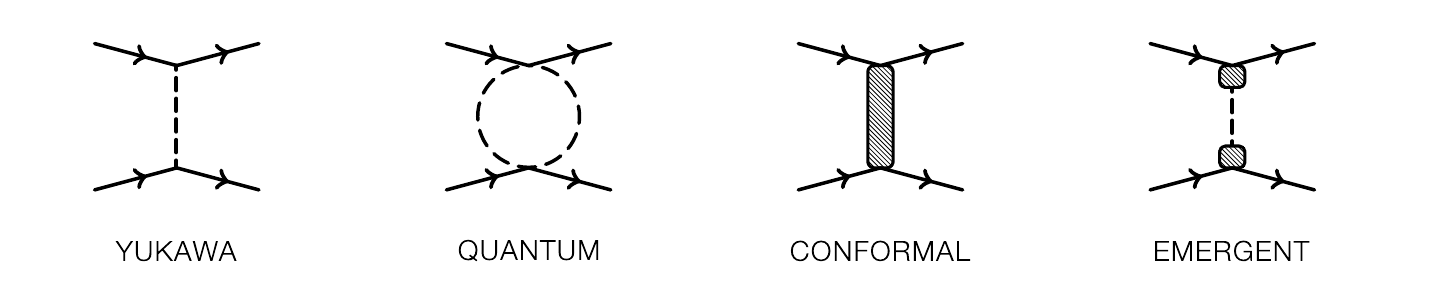}
  \caption{$t$-channel diagrams generating long-range forces in the scenarios we consider. Shaded regions represent strong dynamics.}
  \label{fig:sdforces}
\end{figure}

The complementarity of the exotic and Yukawa-like potentials is manifested clearly in searches for spin-dependent forces.  We point out that spin-dependent Yukawa forces have specific properties and are not representative of the behavior of generic potentials. Because of this, it is necessary to have a set of benchmark scenarios beyond the spin-dependent Yukawa case to interpret and design experiments. The general aspects of effective theory and experimental complementarity are discussed in Sections~\ref{se:properties} and~\ref{se:comp}.
In Sections \ref{sec:quantum}--~\ref{sec:IR} we identify and examine three kinds of exotic potentials based on how they are generated in a microscopic theory:
\begin{itemize}
\item \textit{Quantum}: The  potential is generated by particles that couple bilinearly to nucleons. The leading contribution is generated at loop-level.
\item \textit{Conformal}: The  potential is generated by approximately conformal dynamics.
\item \textit{Emergent}: The  potential is generated by low-mass states that exist in the infrared limit
of the theory, analogous to pions in QCD.
\end{itemize}
These potentials are sketched in Fig.~\ref{fig:sdforces} and can serve as benchmarks for experimental studies. 

\section{Effective Field Theory \label{se:properties}}

We assume that the dark sector interacts  with nucleons via local effective operators.
Because the relevant distance scales are longer than the inverse \acro{QCD} confinement scale, $\Lambda_\text{\acro{QCD}}$, it is sufficient to use a relativistic effective theory of nucleons without specifying the microphysics of dark sector--partonic interactions.

\subsection{Effective Operators and Potentials}

The low-energy effective Lagrangian between nucleons and the dark sector is of the form
\begin{align}
  \mathcal{L} \supset \mathcal{O}_N \mathcal{O}_\text{DS} \ .
  \label{eq:N:DS}
\end{align}
We focus on bilinear nucleon operators $\mathcal O_\text{N} = \bar N \Gamma N$ with some Lorentz structure $\Gamma$. 
We consider the following nucleon bilinears:
\begin{align}
  \mathcal O^\text{S}_N &= \bar N N
  &
  \mathcal O^\text{V}_N &= \bar N \gamma^\mu N
  &
  \mathcal O^\text{T}_N &= \bar N \sigma^{\mu \nu} N 
  \nonumber 
  \\
  \mathcal O^\text{P}_N &= \bar N i\gamma^5 N
  &
  \mathcal O^\text{A}_N &= \bar N \gamma^\mu\gamma^5 N
  \label{eq:ops}
  \ .
\end{align}
We focus on the case where only one of these operators is active in the effective Lagrangian. A pseudo-tensor operator $\bar N  \sigma^{\mu \nu} \gamma^5 N $ also exists, but is redundant with the tensor operator in the scope of our study.~\footnote{This is a consequence of the relation $i \gamma^5 \sigma^{\mu \nu} = \frac{i}{2} \varepsilon^{\mu \nu \alpha \beta} \sigma_{\alpha \beta}$.}
Nucleon bilinears with more complex Lorentz structures also exist but have higher dimension, such that the ones considered here are the most important.~\footnote{
The complexity of the Lorentz structures grows together with the dimensionality of   ${\cal O}_{\rm SM}{\cal O}_{DS}$  because either more derivatives  or a more intricate UV origin---such as higher loop diagrams---are needed to form complex Lorentz structures. 
} 
While the $\mathcal O^{S,P,V,A}$ operators can have renormalizable couplings to spin-0 and spin-1 mediators, the tensor operator has to couple to other fields through a higher-dimensional operator such as a gauge field strength. We remark on ultraviolet completions of this coupling in Appendix~\ref{app:dipole}.

The leading relativistic $2\to 2$ scattering amplitude contributing to long-range potential between nucleons $N_1$ and $N_2$ through operators $\mathcal O^I_N$ and $\mathcal O^J_N$ is
\begin{align}
  i{\cal M}_{IJ} \propto &
  \bra{N_{1,\text{out}}} \mathcal O^I_N \ket{ N_{1,\text{in}} }
  \Sigma_\text{DS}
  \bra{N_{2,\text{out}}} \mathcal O^J_N \ket{ N_{2,\text{in} }} \ 
\end{align}
 where $I\in \{S,V,T,P,A\} $. 
The quantity $\Sigma_\text{DS}$ encodes the intermediate states generated by the dark sector operators $\mathcal O_\text{DS}$. The class of `exotic’ long range potential is encoded in $\Sigma_\text{DS}$. We provide non-relativistic limits of $\bra{N_\text{out}} \mathcal O^I_N \ket{ N_\text{in} }$ in Appendix~\ref{app:NRlims}.

The  potential between nucleons is proportional to the spatial Fourier transform of the  amplitude with respect to the exchanged three-momentum $\vec{q}$, 
\begin{align}
  V_{IJ}(\bm{r}) = -\frac{1}{4m_N^2} \int \frac{d^3 q}{(2\pi)^3} e^{i\vec{q}\cdot \vec{r}}  \mathcal{M}_{IJ} \ . \label{eq:potential:from:amplitude}
\end{align}
The non-relativistic limit is taken by keeping the leading order terms in $|\vec{p}|/m_N$, where $m_N$ is the nucleon mass and $\vec{p}$ is the characteristic nucleon three-momentum.
The cutoff of the effective theory should in principle be taken into account when performing the Fourier integral\footnote{%
In this paper the cutoff, $\Lambda$, is the scale at which the contact operators of the effective relativistic theory are UV completed. The non-relativistic potentials derived from this theory are expansions in $\vec{q}^2/M^2$, where $M$ is typically the mass of the scattering particle. 
For distances shorter than $M^{-1}$, higher order potential terms in $\vec{q}^2/M^2$ are significant when solving the Schr\"odinger equation.
These terms are likely pertinent to resolving the behavior of  singular potentials near the origin~\cite{Kahlhoefer:2017umn}.
Ref.~\cite{Bellazzini:2013foa} takes a complementary, bottom-up approach and addresses singular potentials through a renormalization procedure based on physical observables and assuming only the leading $\vec{q}^2/M^2$ term.
}%
. We show in Appendix~\ref{app:FT_EFT} that effects from the cutoff are negligible when implementing a smooth cutoff avoiding spurious non-analyticity in the integrand.

 In this manuscript we focus on the case where only one operator $\mathcal O^I$ is present so that we examine diagonal potentials $V_I\equiv V_{II}(\bm{r})$. While cross terms $V_{IJ\neq I}(\bm{r})$ may lead to interesting effects, they are necessarily accompanied by the diagonal potentials. Thus in the scope of fifth force searches it is sufficient to focus on diagonal potentials. Table~\ref{fig:big:table}  presents classification of spin structures arising in the potentials studied in this manuscript.

The non-relativistic formalism above takes only $t$-channel diagrams into account and also implies that sources are distinguishable. For certain applications, $u$-channel diagrams can also be relevant see e.g.\ Ref.~\cite{Kahlhoefer:2017umn} for a discussion in the context of self-interacting dark matter.

\subsection{Spin Dependence and Spin Averaging}

Both the amplitude for nucleon scattering, $i \mathcal M$, and the associated long range potential, $V(r)$, are matrices in spin-space.  For instance, the $N_1$ current connects an incoming nucleon spinor to an outgoing nucleon spinor. 
The potential is a tensor product of spin matrices acting on  $N_1$ and $N_2$.  Each component gives the potential for a probe particle of a specified spin scattering off of a source of specified spin. 
The spin-dependent interactions are encoded in linear combinations of Pauli matrices $\bm{\sigma}_{1,2}$ acting on $N_1$ and $N_2$, respectively. The spin-independent pieces are proportional to the $2\times 2$ identities $\mathbbm{1}_1 \otimes \mathbbm{1}_2$. 
The spin structures appearing in the potentials are always tensor products. In this manuscript, we omit the explicit $\otimes$ symbol. The relevant spin structures and their simplified notation are \begin{align}
   \mathbbm{1}_1 \otimes \mathbbm{1}_2   &\equiv \mathbbm{1}_1 \mathbbm{1}_2\,,
   &
  \sigma^i_1 \otimes \sigma^i_2     & \equiv \vec\sigma_1\cdot\vec\sigma_2\,,
   &
   \sigma^i_1 \, \nabla^i \otimes \sigma^j_2 \, \nabla^j  & \equiv 
    (\vec\sigma_1 \cdot \vec\nabla)  (\vec\sigma_2 \cdot \vec\nabla) \equiv \slashed{\nabla}_1 \slashed{\nabla}_2  
      \label{eq:tensor:defs}
\end{align}
where the $1,2$ indices correspond to the $N_1$ and $N_2$ currents.

For unpolarized sources, one averages over  the appropriate initial (or final) state spins in the potential. Observe that this is equivalent to summing together amplitudes with different initial states. This averaging does not change the spin-independent contributions, but causes the spin-dependent contributions to vanish. The scalar and vector potentials are unaffected by polarization average, while the pseudo-scalar and axial potentials vanish at all orders. For the tensor potential a spin-independent contribution remains at next to leading order in the non-relativistic expansion. Spin averaging is denoted by $\langle \sigma_2\rangle$ in Table~\ref{fig:big:table}.

\subsection{Orientation Averaging for Spin-Dependent Potentials}
\label{sec:orientation:avg}

Spin-dependent potentials may be sensitive to scattering orientation. Consider scattering of a probe particle moving along a fixed axis towards the target.  One may obtain the potential for isotropic scattering---for example, in a gas---by appropriate averaging over the polar angle. This is the \textit{orientation-averaged limit}.

Let $\hat{\vec{r}} = \vec{r}/|\vec{r}|$ be the  orientation of source $N_1$ with respect to source $N_2$. 
Certain spin-dependent potentials are proportional to  $\left(\bm{\sigma}_1\cdot \vec\nabla\right)
  \left(\bm{\sigma}_2\cdot \vec\nabla\right)$. Averaging over this orientation  yields
\begin{align}
\left\langle V(\bm{r}) \right\rangle_{\hat{\vec{r}}}
\propto 
\left\langle 
  \left(\bm{\sigma}_1\cdot \vec\nabla\right)
  \left(\bm{\sigma}_2\cdot \vec\nabla\right)\,
  f(r)
\right\rangle_{\hat{\vec{r}}}
  &=
  \frac{1}{3} \left(\bm{\sigma}_1\cdot \bm{\sigma}_2\right)
  \nabla^2 f(r) \ .
  \label{eq:random:orientation}
\end{align}
This particular relation is phenomenologically significant. In the Coulomb case, coming from the exchange of a single massless mediator particle, the radial dependence of the potential is $f(r)\propto 1/r$. Because this is the Green’s function of the three-dimensional Laplacian,
\begin{align}
\left\langle V(\bm{r}) \right\rangle_{\hat{\vec{r}}}
\propto  \nabla^2 f(r) = -4\pi\delta(\bm{r}) \ .
\end{align}
This is simply Gauss’ law. This means that when a single, massless mediator is exchanged, the spin structure in \eqref{eq:random:orientation} produces only a contact interaction and no finite-range contribution. 
Similarly, in case of a Yukawa interaction $f(r)\propto e^{-mr}/r$,   $\nabla^2 f(r)=m^2 f(r) -4\pi\delta(\bm{r}) $ gives a finite range interaction suppressed by $O((mr)^2)$ with respect to the naive dimensional expectation of $1/r^3$.
The suppression of finite-range interactions in the orientation-averaged limit are important for experimental prospects. We discuss this in Section~\ref{se:comp}.  

Orientation averaging is denoted by $\langle \hat r\rangle$ in Table~\ref{fig:big:table}. The columns marked Yukawa have radial dependence characteristic of single particle exchange, whereas the columns marked exotic are general and may be generated by the models presented in this manuscript.

\begin{table}[t]
\centering
\begin{tabular}{@{}llp{.1\textwidth}p{.1\textwidth}p{.1\textwidth}p{.1\textwidth}p{.1\textwidth}p{.1\textwidth}p{.1\textwidth}@{}}
\toprule
&                                          & {
}$V_{S/V}$~[$\mathbbm{1}/\gamma^\mu$] & \multicolumn{2}{c}{$V_P$ ~[$i\gamma^5$]} & \multicolumn{2}{c}{$ V_A$ ~[$\gamma^\mu\gamma^5$]} & 
\multicolumn{2}{c}{$ V_T$ ~[$\sigma_{\mu\nu}$]} \\
      \cmidrule(lr){3-3} \cmidrule(lr){4-5} \cmidrule(lr){6-7} \cmidrule(lr){8-9}
&                                          & \quad {\small Any}                        
& {\small Exotic}                                                       & {\small Yukawa}                                                                 & {\small Exotic}                                                          & {\small Yukawa}                    & {\small Exotic}             & {\small Yukawa} \\
\midrule
\multirow{4}{*}{\rotatebox{90}{\quad\centering \footnotesize no avg.}} 
& $\mathbbm{1}_1\mathbbm{1}_2$                            & \checkmark                                        &                                                                            &                                                                                      &                                                                               &                                         & {\footnotesize $m_N^{-2}r^{-2}$} & {\footnotesize $m^4m_N^{-2}r^2$}\\
& $\slashed{\nabla}_1  \slashed{\nabla}_2$ &                                                   & \checkmark                                                                 & \checkmark                                                                           & \checkmark                                                                    & \checkmark                              & \checkmark                       & \checkmark\\
& $\vec{\sigma_1}\cdot\vec{\sigma_2}$      &                                                   &                                                                            &                                                                                      & \checkmark                                                                    & \checkmark                              & \checkmark                       & {\footnotesize $m^2r^2$}\\
\hline
\multirow{4}{*}{\rotatebox{90}{ \centering \quad  \quad \; \footnotesize $\langle \hat r \rangle$}} 
& $\mathbbm{1}_1\mathbbm{1}_2$                            & \checkmark                                        &                                                                            &                                                                                      &                                                                               &                                         & {\footnotesize $m_N^{-2}r^{-2}$} & {\footnotesize $m^4m_N^{-2}r^{2}$}\\
& $\vec{\sigma_1}\cdot\vec{\sigma_2}$      &                                                   & \checkmark                                                                 & {\footnotesize $m^2 r^2, \delta(r)$ }                                                & \checkmark                                                                    & \checkmark, {\footnotesize $\delta(r)$} &      \checkmark            & {\footnotesize $m^2r^2$} \\
\hline
\multirow{4}{*}{\rotatebox{90}{ \centering \quad \quad \; \footnotesize  $\langle \hat \sigma_2 \rangle$}} 
& $\mathbbm{1}_1\mathbbm{1}_2$                            & \checkmark                                        &                                                                            &                                                                                      &                                                                               &                                         & {\footnotesize $m_N^{-2}r^{-2}$} & {\footnotesize $m^4m_N^{-2}r^{2}$}\\
& $\vec{p}\times\vec{\sigma_1}$            &                                                   &                                                                            &                                                                                      &                                                                     &                                         & {\footnotesize $m_N^{-2}r^{-1}$} & {\footnotesize $m^2m_N^{-2} r$}\\
      \hline
\multirow{4}{*}{\rotatebox{90}{\centering \quad \quad \;\, \footnotesize  $\langle \hat r \rangle$,\,$\langle \hat \sigma_2 \rangle$}} 

& $\mathbbm{1}_1\mathbbm{1}_2$                            & \checkmark                                        &                                                                            &                                                                                      &                                                                               &                                         & {\footnotesize $m_N^{-2}r^{-2}$} & {\footnotesize $m^4m_N^{-2}r^2$} \\ 
& $\vec{p}\times\vec{\sigma_1}$            &                                                   &                                                                            &                                                                                      &                                                                               &                                         &                                  & \\
      \bottomrule
\end{tabular}%

  \caption{Spin structures, \eqref{eq:tensor:defs}, generated by $S,V,T,P,A$ nucleon operators in the cases of no averaging, the orientation-averaged limit, averaging over $N_2$ spins, and both the orientation-averaged limit and averaging over $N_2$ spin. Check marks indicate  that the spin structure is generated. Other factors indicate extra suppression depending on the mediator mass, $m$, and the nucleon mass, $m_N$. Yukawa forces indicate a potential with radial dependence $f(r)\sim e^{-mr}/r$.  Exotic forces indicate a radial dependence that is not Yukawa-like.}
  \label{fig:big:table}
\end{table}

\subsection{Higher-Order Terms}
\label{sec:NLO}

We remark on the contribution of higher-order Feynman diagrams:
\begin{align}
  \vcenter{
    \hbox{\includegraphics[width=.2\textwidth]{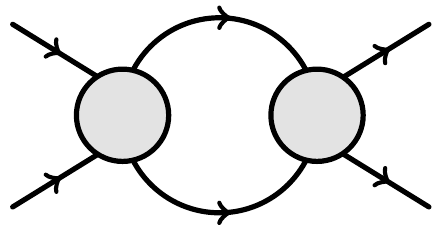}}
    }+\cdots\quad 
    \propto
    \int \frac{d^4k}{(2\pi)^4} 
    \bra{N_{1,\text{out}}} \mathcal O^I_N \Delta_N O^I_N  \ket{ N_{1,\text{in}} }
  \Sigma_\text{DS}
  \bra{N_{2,\text{out}}} \mathcal O^J_N \Delta_N O^J_N \ket{ N_{2,\text{in} }} \ ,
  \label{eq:higher:order}
\end{align}
where $\Delta_N$ is the nucleon propagator. 
The size of such loop diagrams is suppressed compared to tree-level diagrams within the effective theory’s regime of validity.  A good estimate of the magnitude of these diagrams in the non-relativistic limit is obtained by taking $\Delta_N\sim -\mathbbm{1}/m_N$.
However, it is possible that the tree-level diagram may be suppressed by its spin-dependence while the loop diagram gives a spin-independent component. In this case, it is possible that the loop-level contribution produces the most experimentally viable signal. See Section~\ref{se:comp} for more details.

We explicitly illustrate the calculation of a higher order contribution with the case of a pseudoscalar, ${\cal L}=i y \phi \bar N \gamma_5 N$. The tree-level exchange of $\phi$ generates a spin-dependent Yukawa-type force, as indicated in Tab.~\ref{fig:big:table}. At one-loop order, a box diagram made of $\phi$ and $N$ propagators exists. The diagram corresponds to substituting the blobs by tree-level exchange of $\phi$ in  \eqref{eq:higher:order}.\,\footnote{A cross-diagram also exists.} 
The vertices in the amplitude of \eqref{eq:higher:order} are ${\cal O}^{I,J}_N=- y \gamma_5$. 
In  our approximation, we reduce the nucleon propagators $\Delta_N$ to $-\mathbbm{1}/m_N$ in this amplitude. The two $\gamma_5$ simplify, giving  ${\cal O}^{I}\Delta_N{\cal O}^{I}\approx - \mathbbm{1}/m_N\,.
$
One then notices that such contracted diagram is equivalent to the exchange of a  bubble of scalars induced by an effective operator \be  \frac{y^2}{2m_N}  \phi^2 \bar N N \,. \label{eq:eff_phi2} \ee 
Such bubble loop diagram is much simpler to calculate than the original box, and has been evaluated in \textit{e.g.} Ref.~\cite{Brax:2017xho}. The effective vertex \eqref{eq:eff_phi2} gives the spin-independent potential
\be
V(r)=-\frac{y^4}{64\pi^3 m_N^2r^3}\,. \label{eq:VAAAA}
\ee
It turns out this result matches exactly the  result obtained by evaluation of the full box diagram made in \cite{Ferrer:1998ue}. 
Note however that in general our approximation should give  the correct behaviour of the potential but not necessarily reproduce the exact  coefficient.


\section{On Experimental Complementarity \label{se:comp} }

Our effective theory study of the potentials in Section~\ref{se:properties} and the corresponding classification of their properties in Table~\ref{fig:big:table} are useful to evaluate experimental prospects. 
They can be used to sharpen a search strategy and delineate the theoretical motivations of a search.
Whenever an experiment has potential sensitivity to a given type of potential, it is natural to ask whether another effect necessarily exists which may overwhelm the proposed search: either because an experimental bound is already set or because it provides a more promising search direction. In this section, we discuss some examples of experimental complementarity and point out the consequences for existing and future searches. 

Two types of searches for fifth forces are neutron spin rotation and nuclear magnetic resonance experiments.
Neutron spin rotation experiments pass a polarized neutron beam between metal plates that contain unpolarized spins. Spin--velocity potentials cause the polarized neutron spins to precess about the direction of their three-momentum as they pass through the plates. The observation of this precession would be evidence for new physics that generates a spin--velocity potential.
Nuclear magnetic resonance (NMR) experiments can probe new spin-dependent forces through the effective anomalous $J$-coupling of deuterated molecular hydrogen (HD). A gas of HD is in a disordered phase because collisions result in random reorientations of nuclei. The relative orientations between nuclei must be averaged for a fixed separation. This presents a way to search for {orientation-averaged potentials}. We present quantitative details of the experimental reach of NMR experiments in Section~\ref{se:nmr}. 

A simple example of this experimental complementarity is as follows. 
The vector--axial Yukawa cross-potential $V_{VA}$ features a spin-velocity term that can be probed by neutron spin rotation experiments~\cite{Yan:2012wk}. However, any effective theory that yields a $V_{VA}$ potential must simultaneously furnish a vector--vector Yukawa potential $V_V$. This $V_V$ potential is spin-independent and easier to probe experimentally. As a matter of fact, it is highly constrained. In the scope of the search for the dark sector, focusing on $V_V$ is clearly more efficient than $V_{VA}$ so that the search for $V_{VA}$ is therefore not well-motivated.~\footnote{Note that if a new $V_V$ potential were actually discovered, the cross term $V_{VA}$ would then become an efficient way to search for a spin-dependent coupling of the newly discovered particle. }

We see that the complementarity between spin-dependent and spin-independent searches is an important experimental consideration.
To further illustrate this, consider the tree-level exchange of a single pseudoscalar versus the loop-level box diagram coming from the exchange of two pseudoscalars; see Sec.~\ref{sec:NLO}. The former term is spin-dependent while the loop contribution is spin-independent. It is natural to ask which of these contributions is a more effective  channel to search for the new pseudoscalar particle. 
A search for the spin-dependent pseudoscalar potential has been done in \cite{Ledbetter:2012xd} using \acro{NMR} data. 
The experiment is in a disordered phase so that the potentials are calculated in the orientation-averaged limit. This means that the dominant contribution to the tree-level potential vanishes, and the main term is proportional to $ e^{-mr}m^2/(m^2_N r)\ll e^{-mr}/(m^2_N r^3)$. On the other hand, the loop-level contribution force is proportional to $1/(m_N^2 r^3)$ at short distance, as shown in~\eqref{eq:VAAAA}.
The bound on this one-loop, spin-independent component can be readily obtained from \cite{Brax:2017xho}. It turns out that, except in a region of a few orders of magnitude around $m\sim 10$~keV where the \acro{NMR} search is optimal, the bounds from the \textit{spin-independent} component are the dominant ones. There is therefore complementarity between spin-dependent and spin-independent searches, and the spin-independent search turns out to be quite competitive in this example. See \cite{Aldaihan:2016sfj} for another example of this complementarity.

Another key aspect of experimental complementarity is the role of exotic versus Yukawa forces and the corresponding interplay with orientation and spin averages. 
To discuss orientation averaging, consider again the example of a pseudoscalar discussed  above. The Yukawa-like spin-dependent force is suppressed because of Gauss' law. In particular, the force is suppressed by a $\mathcal O\left(m^2 r^2\right)$ factor such that the search sensitivity vanishes in the $m\rightarrow 0$ limit, while the spin-independent component does not. In contrast, an exotic force is---by definition---not subject to cancellations related to Gauss' law, and is therefore not suppressed in the same way. This affects the size of the  spin-dependent force relative to higher-order spin-independent counterparts. 
Moreover, in contrast to scenarios with Yukawa-like forces, one typically expects that the higher-loop contributions to exotic forces are subdominant and poorly constrained. 
From this, one concludes that  exotic forces motivate experiments with disordered phases---such as \acro{NMR}-based searches~\cite{Ledbetter:2012xd}. 

The case of a single polarized source is also interesting. Table~\ref{fig:big:table} shows that the tensor potential features a mixed velocity--spin structure, $\vec{p}\times\vec\sigma_1$, in this limit. 
The complementarity of Yukawa versus exotic forces is again relevant in this case.
For a Yukawa-type force, the velocity-dependent structure is $\mathcal O(m^2)$ and thus vanishes in the $m\rightarrow 0$ limit. This is not the case for an exotic force. 
One concludes that exotic forces motivate searches for velocity--spin dependent forces. These can be done in experiments with a polarized source---such as a neutron beam---at finite velocity.
We expand on the theoretical ultraviolet completion of the tensor operator in Appendix~\ref{app:dipole}.

In summary, the possible existence of exotic spin-dependent forces motivates experiments with (i) disordered phases and (ii) polarized sources at finite velocity.
Another general lesson is that the properties of Yukawa-like forces are rather non-generic and so they should not be used as the only benchmark for spin-dependent fifth force experiments. The following sections present well-motivated alternatives that can serve as benchmark models for the exotic potentials.

\section{Spin-Dependent Quantum Forces}
\label{sec:quantum} 

Our first example of an exotic force is when $\mathcal O_\text{DS}$ in \eqref{eq:N:DS} is a bilinear of dark particles, each with mass $m$. The potential is generated by the exchange of \emph{two} virtual particles~\cite{Feinberg:1968zz}. In the scenario where the dark particles are dark matter, then the bilinear interaction with nucleons may be motivated from a $\mathbb{Z}_2$ symmetry that explains the particle’s stability.\,\footnote{The search for long-range potentials induced by the loop-level exchange of pairs of dark particles is then a search for virtual dark matter. In contrast to searches for on-shell dark matter in direct detection experiments, this is independent of the dark matter phase space distribution in the local galaxy.} 

We extend the spin-independent study in Ref.~\cite{Fichet:2017bng} to the case of spin-dependent nucleon operators. 
We write the combined $\mathcal O_N \cal O_\text{DS}$ operators in  \eqref{eq:N:DS} as effective contact operators.
Because $\mathcal O_\text{DS}$ is a dark particle bilinear, it is possible to have dark particles of any spin. Here we consider spin-0, $\frac 12$, and 1.
These operators are non-renormalizable so that the effective theory is valid above distance scales of
\begin{align}
  r
    \sim
  \text{max}\left(\frac{1}{4\pi \Lambda}\, , \, \frac{1}{4\pi\Lambda_\text{QCD}} \right) \ ,
\end{align}
where $\Lambda$ is the scale at which the contact operator description breaks down. This is defined by the underlying physics that generate the $\mathcal O_\text{N} \cal O_\text{DS}$ operators.
We consider the operators:
\begin{align}
     \mathcal{O}_a^0&=\frac{1}{\Lambda^2}\bar{N}\gamma^\mu \gamma^5 N i \phi^* \overleftrightarrow{\partial_\mu} \phi & 
     \mathcal{O}_a^\frac{1}{2}&=\frac{1}{\Lambda^2}\bar{N}i\gamma^5 N \bar{\chi}i\gamma^5 \chi \nonumber & \mathcal{O}_a^1&=\frac{1}{\Lambda^3}\bar{N}i\gamma^5 N |X_{\mu\nu}|^2\nonumber\\
     \mathcal{O}_b^0&=\frac{1}{\Lambda}\bar{N}i\gamma^5 N |\phi|^2 & 
     \mathcal{O}_b^\frac{1}{2}&=\frac{1}{\Lambda^2}\bar{N}i\gamma^5 N \bar{\chi}\chi \nonumber & \mathcal{O}_b^1&=\frac{1}{\Lambda^3}\bar{N}i\gamma^5 N X_{\mu\nu}\tilde{X}^{\mu\nu}\nonumber\\
     \mathcal{O}_c^0&=\frac{1}{\Lambda^3}\bar{N}i\gamma^5 N|\partial_\mu\phi|^2 & 
     \mathcal{O}_c^\frac{1}{2}&=\frac{1}{\Lambda^2}\bar{N}\gamma^\mu\gamma^5 N \bar{\chi}\gamma_\mu\gamma^5 \chi \\
     \mathcal{O}_d^0&=\frac{1}{\Lambda^3}\bar{N}\sigma^{\mu\nu} N(\partial_\mu\phi)(\partial_\nu\phi^*) & 
      \mathcal{O}_d^\frac{1}{2}&=\frac{1}{\Lambda^2}\bar{N}\gamma^\mu\gamma^5 N \bar{\chi}\gamma_\mu \chi \nonumber  \nonumber\\  & &
     \mathcal{O}_e^\frac{1}{2}&=\frac{1}{\Lambda^2}\bar{N}\sigma^{\mu\nu} N \bar{\chi}\sigma_{\mu\nu} \chi & & &.\nonumber
     \label{eq:loopy:operators}
\end{align} 
We write the dark particle as $\phi$ for scalars, $\chi$ for  fermions, and $X^\mu$ for vectors. We use  $\overleftrightarrow{\partial_\mu}\equiv\overrightarrow{\partial_\mu} - \overleftarrow{\partial_\mu}$. 
We assume that only one of these operators is active. That is, we do not consider any cases that mix vertices from different operators in a single Feynman diagram.

The potentials depend on modified Bessel functions of the second kind and on a Meijer $G$-function; we denote these by:
\begin{align}
K_i&\equiv K_i(2mr)
&
  G&\equiv G_{2,4}^{2,0}\left(m^2 r^2\bigg|
  \begin{array}{c}
  \frac{1}{2},\frac{3}{2} \\
  0,0,\frac{1}{2},\frac{1}{2} \\
  \end{array}
  \right) \ .
\end{align}
We also introduce a discrete variable for whether or not the dark particle is self-conjugate:
\begin{align}
  \eta &= 
  \begin{cases}
  0 & \text{if self-conjugate}\\
  1 & \text{otherwise}
  \end{cases}
  \ .
\end{align}

For these loop-induced potentials, the amplitude has a branch cut---appearing via a logarithm. The Fourier transform integral \eqref{eq:potential:from:amplitude} 
 is performed by analytical continuation into the complex $|\vec{q}|$ plane, reducing the integral  over the real line to one on the discontinuity across the branch cut. 
 Details are provided in Appendix~\ref{app:pots}. 

The operators produce the following potentials:
\begin{align}
  V_a^0
    &=
    -\eta\frac{m^2(\vec{\sigma_1}\cdot\vec{\sigma_2})}{8\pi^3\Lambda^4}\left(\frac{K_2}{r^3}\right)
    -\eta\frac{(\vec{\sigma_1}\cdot \vec{\nabla})(\vec{\sigma_2}\cdot \vec{\nabla})}{96\pi^3\Lambda^4}\left(\frac{-2mrK_1+4+m^2\pi^2r^2G}{r^3}\right)
    \\
  V_b^0
  &=
    2^\eta\frac{m(\vec{\sigma_1}\cdot \vec{\nabla})(\vec{\sigma_2}\cdot \vec{\nabla})}{128\pi^3\Lambda^2m_N^2}\left(\frac{K_1}{ r^2}\right)
  \\
  V_c^0
  &=
  2^\eta\frac{m^2(\vec{\sigma_1}\cdot \vec{\nabla})(\vec{\sigma_2}\cdot \vec{\nabla})}{128\pi^3\Lambda^6m_N^2}\left(\frac{(15mr+m^3r^3)K_1+(30+6m^2r^2)K_2}{r^5}\right)
  \\
  V_d^0 
  &=
  \eta\frac{\left[(\vec{\sigma_1}\cdot \vec{\nabla})(\vec{\sigma_2}\cdot \vec{\nabla})-(\vec{\sigma_1}\cdot\vec{\sigma_2})\nabla^2\right]}{32\pi^3\Lambda^6}\left(\frac{m^2K_2}{r^3}\right) \\ V_a^\frac{1}{2}
&=
2^{\eta}\frac{(\vec{\sigma_1}\cdot \vec{\nabla})(\vec{\sigma_2}\cdot \vec{\nabla})}{32\pi^3m_N^2\Lambda^4}
\left(\frac{3m^2K_2-2m^3rK_1}{r^3} \right)
\\
V_b^\frac{1}{2}
&=
2^{\eta}\frac{(\vec{\sigma_1}\cdot \vec{\nabla})(\vec{\sigma_2}\cdot \vec{\nabla})}{32\pi^3m_N^2\Lambda^4}
\left(\frac{3m^2K_2}{ r^3}\right)
\\
V_c^\frac{1}{2}
&=
-2^{\eta}\frac{m^2(\vec{\sigma_1}\cdot\vec{\sigma_2})}{4\pi^3\Lambda^4}
\left(\frac{K_2}{r^3}\right)
+
2^{\eta}\frac{(\vec{\sigma_1}\cdot \vec{\nabla})(\vec{\sigma_2}\cdot \vec{\nabla})}{96 \pi^3 \Lambda^4 }
\left(\frac{\pi ^2 m^2 r^2 G+4 m r K_1+4}{r^3}\right)
\\
V_d^\frac{1}{2} 
&=
-\eta\frac{m^2(\vec{\sigma_1}\cdot\vec{\sigma_2})}{2\pi^3\Lambda^4}
\left(\frac{mrK_1+K_2}{r^3}\right)
+
\eta\frac{(\vec{\sigma_1}\cdot \vec{\nabla})(\vec{\sigma_2}\cdot \vec{\nabla})}{48 \pi^3 \Lambda^4 }
\left(\frac{\pi ^2 m^2 r^2 G+4 m r K_1+4}{r^3}\right)
\\
V_e^\frac{1}{2} 
&=
-\eta\frac{ m^2 (\vec{\sigma_1}\cdot\vec{\sigma_2})}{ \pi^3\Lambda^4}
\left(\frac{K_2}{r^3}\right)
-
\eta\frac{\left[(\vec{\sigma_1}\cdot \vec{\nabla})(\vec{\sigma_2}\cdot \vec{\nabla})-(\vec{\sigma_1}\cdot\vec{\sigma_2})\nabla^2\right]}{12 \pi^3\Lambda^4}
\left(\frac{\pi ^2 m^2 r^2 G+4 m r K_1+4}{r^3}\right)\\ V_a^1 
&=
2^{\eta}\frac{(\vec{\sigma_1}\cdot \vec{\nabla})(\vec{\sigma_2}\cdot \vec{\nabla})}{8\pi^3 m_N^2\Lambda^6}\left[\left(\frac{30m^3}{r^4}+\frac{3m^5}{r^2}\right)K_1+\left(\frac{60m^2}{r^5}+\frac{12m^4}{r^3}\right)K_2\right]
\\
V_b^1 
&=
2^{\eta}\frac{(\vec{\sigma_1}\cdot \vec{\nabla})(\vec{\sigma_2}\cdot \vec{\nabla})}{8\pi^3 m_N^2\Lambda^6}
\left(\frac{30m^3}{r^4}K_3+\frac{12m^4}{r^3}K_2\right)\ . \label{eq:V1b}
\end{align}

\subsection{Orientation-Averaged Limit}

We perform the orientation-averaging limit of Section~\ref{sec:orientation:avg} on the potentials of the previous section. 
We present these in the $r\ll m^{-1}$ limit. Note that by doing so, the mass dependence drops out. These take the limiting forms:

\begin{align}
    V_a^0&=-\eta\frac{\vec{\sigma_1}\cdot\vec{\sigma_2}}{24\pi^3\Lambda^4}\frac{1}{r^5} &
    V_a^\frac{1}{2}&=2^{\eta}\frac{\vec{\sigma_1}\cdot \vec{\sigma_2}}{16\pi^3m_N^2\Lambda^4}\frac{5}{ r^7} \nonumber & 
    V_a^1 &= 2^{\eta}\frac{\vec{\sigma_1}\cdot \vec{\sigma_2}}{2\pi^3 m_N^2\Lambda^6}\frac{105}{r^9}\nonumber\\
  V_b^0&=2^\eta\frac{\vec{\sigma_1}\cdot \vec{\sigma_2}}{128\pi^3m_N^2\Lambda^2}\frac{1}{r^5} &
    V_b^\frac{1}{2}&=2^{\eta}\frac{\vec{\sigma_1}\cdot \vec{\sigma_2}}{16\pi^3m_N^2\Lambda^4}\frac{5}{ r^7} \nonumber & 
    V_b^1 &= 2^{\eta}\frac{\vec{\sigma_1}\cdot \vec{\sigma_2}}{2\pi^3 m_N^2\Lambda^6}\frac{105}{r^9} \nonumber\\
    V_c^0&=2^\eta\frac{\vec{\sigma_1}\cdot\vec{\sigma_2}}{64\pi^3m_N^2\Lambda^6}\frac{105}{r^9} & 
    V_c^\frac{1}{2}&=-2^{\eta}\frac{\vec{\sigma_1}\cdot\vec{\sigma_2}}{12\pi^3\Lambda^4}\frac{1}{r^5}  \\
    V_d^0 &=-\eta\frac{\vec{\sigma_1}\cdot\vec{\sigma_2}}{24\pi^3\Lambda^6}\frac{5}{r^7} &
    V_d^\frac{1}{2} &= -\eta\frac{\vec{\sigma_1}\cdot\vec{\sigma_2}}{6\pi^3\Lambda^4}\frac{1}{r^5} & \nonumber\\  & & V_e^\frac{1}{2} & = \eta\frac{\vec{\sigma_1}\cdot\vec{\sigma_2}}{6\pi^3\Lambda^4}\frac{1}{r^5} & & & .\nonumber
  \end{align}
  
  \subsubsection{Bounds on Dark Matter from NMR\label{se:nmr}}
  
  \begin{figure}
			\centering
 			\includegraphics[width=0.55\linewidth]{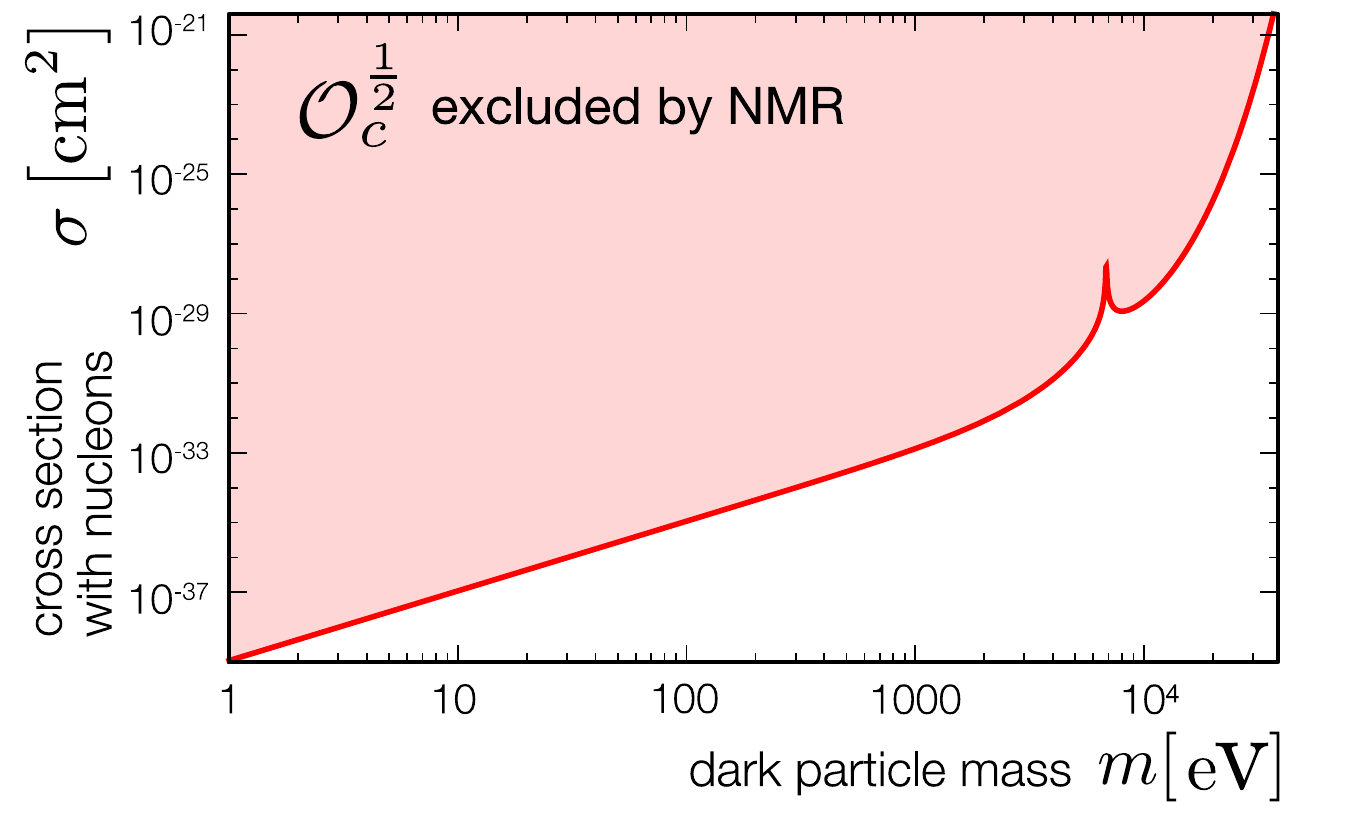}
			\caption{Bounds on $\mathcal{O}_c^\frac{1}{2}$ from NMR~\cite{Ledbetter:2012xd} on the direct detection plane. Note that this search is sensitive to much lighter masses than direct detection experiments, e.g.~\cite{Aprile:2019dbj}. 
			The cross section is for tree-level $2\rightarrow2$ scattering of dark matter off xenon. The bound vanishes in a narrow region around $m\approx 6850$~eV because orientation averaging causes $V_c^\frac{1}{2}$ to vanish at approximately $2mr\approx 5.2$.}
			\label{fig:nmr-sd-bounds}
		\end{figure}

The orientation-averaged form of the potential appears when forces are present in a disordered phase of matter. This is the case for NMR experiments in a gas phase. NMR data can be used to place bounds on the effective anomalous J-coupling of  deuterated molecular hydrogen  HD; Ref.~\cite{Ledbetter:2012xd} found that $\Delta J_3$ must be less than $9.8 \times 10^{-16}$ eV.  This, in turn, bounds the strength of a possible spin-dependent potential between these nuclei.
The proton-deuteron distance is of order  $\langle r \rangle=0.00038 $\,eV$^{-1}$. The relative orientations between nuclei are averaged because of  random reorientation of HD molecules due to collisions. 

If one of the dark particles $\phi$, $\chi$ or $X^\mu$ is identified as the dark matter, the contact interactions \eqref{eq:loopy:operators} induce dark matter--nucleus scattering. This scattering process is probed 
by dark matter direct detection experiments.
The sensitivity of such searches typically drops below $m\sim 1$\,GeV. In contrast, NMR experiments probe dark particles that are much lighter than this scale---illustrating the  complementarity between  dark matter scattering and quantum force searches.

As a specific example, we provide bounds on Dirac  dark matter interacting with xenon through the $\mathcal{O}_c^\frac{1}{2}$  axial-vector interaction. This is the standard benchmark case for spin-dependent direct detection. The bound is plotted in Fig.~\ref{fig:nmr-sd-bounds}.

\subsection{Spin-dependent Potentials with One Unpolarized Source}

In the case in which one source is unpolarized, one must average over source's initial state spins. The corresponding nucleon bilinear vanishes for pseudoscalar ($N i\gamma_5 N$) and  axial vector ($\bar N \gamma_\mu\gamma_5 N$) interactions. 
This leaves only the operators with spin-0 or spin-1/2 dark particles interacting with the nucleon dipole. 
The resulting potentials include a term that is spin independent term and a term that is both spin and velocity dependent. 
We define the velocity, $\vec v$ as the average of the probe nucleon’s incoming and outgoing momenta divided by the nucleon mass.

The general long-range potentials are:
\begin{align}
  V_d^0 
&=
\eta 
\frac{%
  2m_N\mathbbm{1}_2(\vec{v}\times \vec{\sigma_1})\cdot\vec{\nabla}+ \mathbbm{1}_1 \mathbbm{1}_2\nabla^2%
}{%
  128\pi^3m_N^2\Lambda^6
}
\nabla^2
\left(
  \frac{m^2K_2}{r^3}
\right) \\ V_e^\frac{1}{2}
&=
\eta
\frac{%
  2m_N \mathbbm{1}_2(\vec{v}\times \vec{\sigma_1})\cdot\vec{\nabla}
  + \mathbbm{1}_1 \mathbbm{1}_2\nabla^2%
}{%
4\pi^3 m_N^2\Lambda^4
}
m^2
\left(\frac{2mrK_1+K_2}{r^3}\right) \ .
\end{align}

At distances much smaller than the dark particle's Compton wavelength, $m^{-1}$, these reduce to:
\begin{align}
  V_d^0(r\ll m^{-1})  
&=
\eta
\frac{5}{64\pi^3m_N^2\Lambda^6}
\left(
  \frac{42\mathbbm{1}_1 \mathbbm{1}_2}{r^9}
  +2m_N\mathbbm{1}_2
    (
    \vec{v} 
    \times \vec{\sigma_1})
    \cdot\vec{\nabla}
  \frac{1}{r^7}
\right)
\\ V_e^\frac{1}{2}(r\ll m^{-1})  
&= 
\eta
\frac{1}{8\pi^3 m_N^2\Lambda^4}
\left(
  \frac{20\mathbbm{1}_1 \mathbbm{1}_2}{r^7}
  +2m_N\mathbbm{1}_2
    (
    \vec{v}
    \times \vec{\sigma_1})
    \cdot\vec{\nabla}
  \frac{1}{r^5}
\right)\ .
\end{align}

\section{Spin-Dependent Warped/Conformal Forces}
\label{sec:warped}

Weakly coupled new physics in four dimensions produce potentials that carry  negative integer powers of $r$. 
This section presents a departure from this behavior by examining effects from a five dimensional curved space and from a four-dimensional conformal sector.

\subsection{Warped Dark Sector Scenario}

The Standard Model may live on a four-dimensional brane lying at the boundary of a truncated five-dimensional AdS space with curvature $k$; see {e.g.} \cite{ArkaniHamed:2000ds, Randall:1999vf, Brax:2003fv}. 
We refer to this brane as the \acro{UV} brane.  
In this model, the effect of 5D gravity is mild as it is localized away from the \acro{UV} brane. 5D gravity only induces a small $\mathcal O(k^{-2}r^{-3})$ correction to Newton potential. Such deviation is mildly constrained experimentally:  torsion pendulum experiments put a lower bound $k\gtrsim 10^4 $~TeV, which constitutes the leading constraint for curvature in this model~\cite{Brax:2017xho}. On the other hand, the AdS curvature $k$ may be as high as the 4D Planck mass while staying consistent with the validity of the 5D effective theory.

\begin{figure}
  \centering
  \includegraphics[height=.3\textwidth]{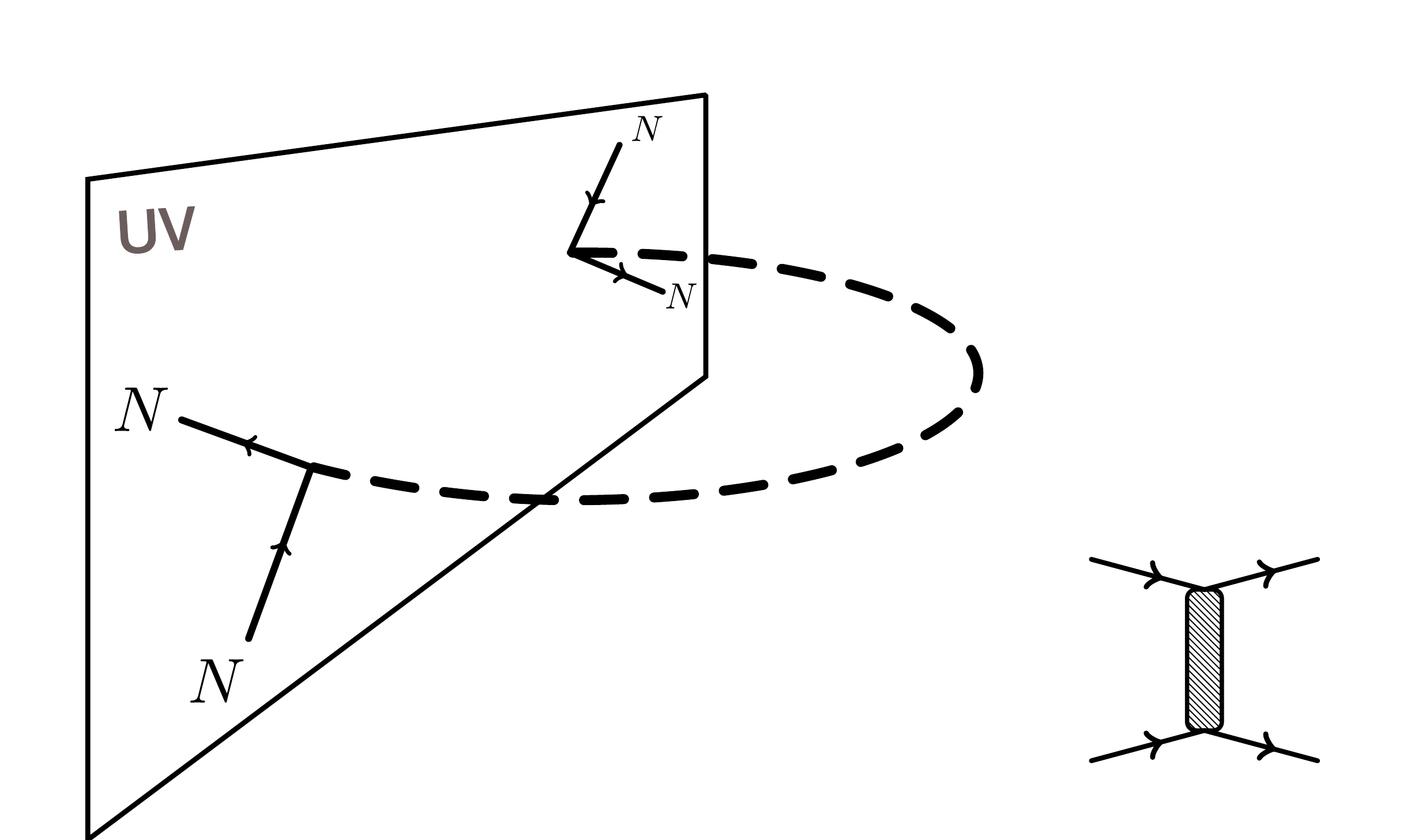}
  \caption{Diagram in 5D anti-de Sitter space giving rise to the conformal force.}
  \label{fig:conformal}
\end{figure}

Beyond this minimal braneworld model, matter fields can in principle propagate in the bulk of the extra dimension. These fields tend to be hidden from the \acro{UV} brane as a result of the localization in the fifth dimension induced by the curvature. This framework therefore naturally gives rise to a dark sector~\cite{Brax:2019koq} and is sketched in Fig.~\ref{fig:conformal}.
As a concrete realization, consider the simplest case of a bulk scalar field, $\Phi$ that couples linearly to nucleons. The 5D matter action is
\begin{align}
S_5
\supset 
\int_{\rm bulk} d^5X \sqrt{|g|}
\left(  
  \frac{1}{2}\partial_M \Phi \partial^M \Phi -\frac{m^2_\Phi}{2} \Phi^2 \right) 
+
\int_{\rm brane} d^4X  \sqrt{|\gamma|}
\left( 
  {\cal L}_{\rm SM} +\frac{\lambda}{\sqrt{k}}{\cal O}_\text{N} \Phi    
  - 
 \frac{m_{\rm UV}}{2} \Phi^2
\right) 
\,,  
\label{eq:warped:S5}
\end{align}
where $\gamma_{\mu\nu}$ is the induced metric on the brane, $\mathcal O_N$ is a nucleon bilinear, and $\lambda$ is a dimensionless effective coupling which can be taken to be of order one. 
Using conformal coordinates and a $(+,-,-,-,-)$ signature, the AdS metric reads
\begin{align}
ds^2=g_{MN}X^M X^N= \frac{1}{(kz)^2}\left(\eta_{\mu\nu}dx^\mu dx^\nu -dz^2  \right) \ ,
\end{align}
and the brane is localized at
$z_0=k^{-1}$.\,\footnote{Another common convention in the literature is to use the curvature radius of AdS, $R\equiv k^{-1}$.} We focus on the case of a brane-localized mass term satisfying the condition $m_{\rm UV}=
\sqrt{4k^2+m^2_\Phi} - 2k \,.
$ This is consistent with the \acro{BPS} brane condition from supergravity, and can be also motivated using holography, see \cite{Brax:2003fv} and references therein.

We consider couplings of the bulk scalar to the scalar ($\mathcal O_S$) and pseudoscalar ($\mathcal O_P$) nucleon operators in \eqref{eq:ops}.
The former generates a spin-independent force which we include as a useful benchmark.
The axial coupling $\frac{1}{\Lambda}\bar N \gamma_5 \gamma_\mu N \partial_\mu \Phi $ generates the same tree-level potential as the
pseudoscalar. This can be seen directly at the level of the Lagrangian by integrating by part the interaction and using the nucleon equation of motion. \footnote{The theories differ at higher order in their interactions.
At loop level, for example, the spin-independent potential from a scalar with a derivative coupling to an axial vector current can be distinguished from a pseudoscalar~\cite{Ferrer:1998ue}.
}

\subsection{Potential}

The brane-to-brane Feynman propagator for $\Phi$ in mixed position/momentum space~\cite{Puchwein:2003jq, Carena:2003fx} reads
\begin{align}
  \langle \Phi(p^\mu,z_0) \Phi(-p^\mu,z_0) \rangle
  &\equiv 
  \Delta_p(z_0,z_0) 
  =\frac{i}{p}\frac{H_\alpha^{(2)}(p/k)}{H_{\alpha-1}^{(2)}(p/k)} 
  &
  p&= \sqrt{\eta_{\mu \nu} p^\mu p^\nu}
  &
  \alpha &= \sqrt{4+\frac{m_\Phi^2}{k^2}} 
\, ,
\end{align}
where $H^{(2)}$ is the Hankel function of the second kind. 
Regularity of the solution to the 5D equation of motion is imposed to obtain this propagator. 
For invariant four-momenta much less than the curvature, $p\ll k$, this gives
\begin{align}
  \Delta_p(z_0,z_0) 
  \approx 
  i\left[\frac{p^2}{2k (\alpha-1)}   - 2 k\, \frac{\Gamma(1-\alpha)}{\Gamma(\alpha)}\,\left(-\frac{p}{2k}\right)^{2\alpha} \right]^{-1} \,. 
  \label{eq:AdSfull_prop}
\end{align}

For $\alpha>1$, this propagator is dominated by the analytic term $\propto 1/p^2$.  
An observer on the \acro{UV} brane  thus mostly see a 4D massless field, similarly to the gravity case discussed above.
For $\alpha<1$, on the other hand, the propagator \eqref{eq:AdSfull_prop} is  dominated by the non-analytic term with $p^{2\alpha}$ scaling.

One may qualitatively understand this behavior by imagining the presence of an \acro{IR} brane at some distance $z_1 > z_0$ that creates a discrete spectrum of Kaluza--Klein modes.
The parameter $\alpha$ controls the localization of these modes.
 For $\alpha>1$ and any boundary condition on the \acro{IR} brane,  an ultralight mode localized on the \acro{UV} brane exists. This corresponds to the 4D pole in $\Delta_p(z_0,z_0)$.
 For $\alpha<1$, the light mode is instead localized towards the \acro{IR} brane and an observer on the \acro{UV} brane is primarily sensitive to the Kaluza--Klein excitations that are collectively encoded in the non-analytic $p^{2\alpha}$ term.

For the case $\alpha<1$, the bulk scalar $\Phi$ generates an exotic long-range force between the nucleons on the \acro{UV} brane. This force has \emph{non-integer} dependence on the nucleon separation. 
The scalar and pseudo-scalar nucleon operators in \eqref{eq:ops} generate the following potentials:
\begin{align}
  V_S(r) 
  &= 
  \frac{-\lambda^2}{2\pi^\frac{3}{2}}
  \frac{\Gamma(\frac{3}{2}-\alpha)}{\Gamma(1-\alpha)}
  \frac{1}{r(kr)^{2-2\alpha}} 
  \label{eq:warped:V:s}
  \\
  V_P(r) 
  &= \frac{\lambda^2}{2\pi^\frac{3}{2}}
  \frac{\Gamma(\frac{3}{2}-\alpha)}{\Gamma(1-\alpha)}
  \frac{(\vec{\sigma_1}\cdot \vec{\nabla})(\vec{\sigma_2}\cdot \vec{\nabla})}{4m_N^2}\left(\frac{1}{r(kr)^{2-2\alpha}}\right) 
  \label{eq:warped:V:p}
  \,. 
\end{align}

In the orientation-averaged limit, the pseudo-scalar potential becomes:
\begin{align}
V_P(r) 
  &= \frac{2\lambda^2}{3\pi^\frac{3}{2}}
  \frac{(1-\alpha)\Gamma(\frac{5}{2}-\alpha)}{\Gamma(1-\alpha)}
  \frac{(\vec{\sigma_1}\cdot\vec{\sigma_2})}{4m_N^2}\frac{1}{r^3(kr)^{2-2\alpha}}.
\end{align}

\subsection{A Conformal Model}
\label{sec:warped:conformal}

The AdS/CFT correspondence relates the warped dark sector model to a purely four-dimensional conformal dark sector.
For an exotic potential with $\alpha<1$, the simplest realization of the duality is 
\begin{align}
  {\mathcal L}={\mathcal L}_{\rm SM}+{\mathcal L}_{\rm CFT} + \frac{1}{M^{\Delta_{\rm SM}+\Delta_{\rm CFT}-4}}{\cal O}_{\rm N}{\cal O}_{\rm CFT}\,. 
  \label{eq:LCFT}
\end{align}
The Lagrangian ${\mathcal L}_{\rm CFT}$ encodes the dynamics of the conformal field theory. The operator ${\mathcal O}_{\rm CFT}$ contains fields from the conformal sector. $\Delta_{\rm SM}=3$ is the scaling dimension of the scalar or pseudo-scalar nucleon operators.
We further define $\Delta \equiv \Delta_{\rm CFT}$, the dimension of the ${\cal O}_{\rm CFT}$ operator.  

The symmetries of the \acro{CFT} fully dictates the theory’s behavior.
In position space, conformal symmetry constrains the two-point correlation function to be 
\begin{align}
 \langle{\cal O}_{\rm CFT}(0)\,{\cal O}_{\rm CFT}(x)\rangle=\frac{c}{4\pi^2(-|x|)^{2\Delta}}   \ ,
\end{align}
where $c$ is an undetermined real number. In momentum space this is
\begin{align}
  \langle{\cal O}_{\rm CFT}(-p)\, {\cal O}_{\rm CFT}(p)\rangle
  &=
  c\frac{i\, \Gamma(2-\Delta)}{4\,\Gamma(\Delta)}\left(\frac{-p^2}{4}\right)^{\Delta-2}  \,.
\label{eq:CFT_corr}
\end{align}
By construction, nucleons source this correlation function and thus experience a force due to the conformal dynamics.

By comparing \eqref{eq:CFT_corr} to the non-analytical part of the brane-to-brane propagator in \eqref{eq:AdSfull_prop}, one identifies $\Delta=2-\alpha$. 
In the language of {AdS}/\acro{CFT}, this is the ``$\Delta_-$ branch'' of the correspondence, in which the boundary of the bulk field is identified with the classical value of ${\cal O}_{\rm CFT}$ itself.
This version of the duality is valid only for $0<\alpha<1$, {i.e.} $1<\Delta<2$. In contrast, the $\Delta_+$ branch identifies $\Delta = 2+\alpha$ and is valid for any $\alpha$. The $\Delta_-$ branch model is sufficient in the context of our study.

Taking the non-relativistic limit of the conformal two-point functions and transforming back to position space yields the \emph{conformal} spin-dependent potentials analogous to~(\ref{eq:warped:V:s}--\ref{eq:warped:V:p}): 
\begin{align}
  V_S(r) &= 
  \frac{c}{4\pi^{3/2}}
  \frac{\Gamma(\Delta-1/2)}{\Gamma(\Delta)}
  \frac{1}{r(Mr)^{2\Delta-2}}
  \\
  V_P(r) &= 
  -\frac{c}{4\pi^{3/2}}
  \frac{\Gamma(\Delta-1/2)}{\Gamma(\Delta)}
  \frac{(\vec{\sigma_1}\cdot
  \vec{\nabla})(\vec{\sigma_2}\cdot \vec{\nabla})}{4m_N^2}
  \left(\frac{1}{r(Mr)^{2\Delta-2}}\right) \ .
\end{align}
Both the warped and conformal models continuously interpolates between $1/r$ and $1/r^3$ as a function of a continuous parameter: either the bulk mass of the scalar field living AdS space, or---equivalently---the conformal dimension of the correlation function of conformal dynamics exchanged between nucleons.

\section{Spin-Dependent Emergent Forces}
\label{sec:IR}

A new force may emerge from a  change in a theory’s degrees of freedom at low energies. This may happen, for example, as a consequence of a phase transition. This scenario departs from Yukawa-like behavior and is thus exotic. Also, there is a priori no principle enforcing exact continuity of the potential between the short- and long-distance regimes, hence the  potential  might feature a smooth kink in the transition regime.  We discuss this scenario from the perspective of a confining non-Abelian gauge theory and present a more explicit realization  from a five-dimensional holographic setup.

\subsection{Emergent Force from a Confining Dark Sector \label{se:nonab}}

Suppose a dark sector contains a non-Abelian gauge field with coupling $g$ and field strength $X_{\mu \nu}$. This sector can couple to visible matter through gauge-invariant higher-dimensional operators of the form \eqref{eq:N:DS}:
\begin{align}
  \mathcal L 
  \supset 
  \frac{c}{M^3} \mathcal O_\text{N} X_{\mu \nu}X^{\mu \nu}
  + \cdots \ .
\end{align}
In the regime where the dark sector gauge theory is weakly coupled, it induces a quantum force
(see Section~\ref{sec:quantum}), whose exact form is given by \eqref{eq:V1b}. 
Dimensional analysis  requires the potential to scale as
\begin{align}
  V(r)_\text{gauge} \sim \frac{c^2 \,g^4}{M^6 r^7} \ .
    \label{eq:emergent:deconfined}
\end{align}
When the gauge theory is asymptotically free, there is an infrared scale, $\mu \ll M $, at which the theory becomes strongly-interacting and confines. In this scenario, the theory may develop a mass gap analogous to \acro{QCD}. 
Below the confinement scale, physics is described by an effective theory of composite states.  
Composite states with same quantum numbers as $ \text{Tr } X_{\mu\nu} X^{\mu\nu}$ may arise as scalar glueball fields:
\begin{align}
  \mathcal{L}_\text{conf}
  \supset
  \frac{c_\varphi \mu^3}{M^3}
  {\cal O}_{\rm SM}\, \varphi + \ldots \,,
\end{align}
where $\varphi$ is a CP-even scalar denoting the glueball excitation in the confined theory.  
See Refs.~\cite{Boddy:2014qxa,Boddy:2014yra} for earlier work on dark sectors described by glueball-like excitations. The $c_\varphi$ coefficient arises from strong dynamics and is evaluated by dimensional analysis at strong coupling.

In the confined phase, $r \gg \mu^{-1}$, the potential is Yukawa-like, 
\begin{align}
  V_\text{conf}(r)
  \sim
  \frac{c^2_\varphi}{r}\frac{\mu^6}{M^6}
  \,.
  \label{eq:emergent:confined}
\end{align}
The potentials \eqref{eq:emergent:deconfined} and \eqref{eq:emergent:confined} appear to match at the level of dimensional analysis: the transition occurs at $r\sim 1/\mu$, where it turns out that $V_{\rm conf}(\mu^{-1})\sim V_{\rm gauge}(\mu^{-1})$. 
However, the $c^2g^4$ and $c_\varphi^2$ couplings are in principle different. 
The coupling $g$ is perturbative at energies well above the confinement scale $\mu$. In contrast, $c_\varphi$ may typically be $\mathcal O(4\pi)$, hence 
it is possible that $c_\varphi^2$ be substantially larger than $c^2g^4$. This implies that the spatial potential could undergo some {enhancement} when $r$  enters the confined region at  $r\sim 1/\mu$.

\subsection{Holographic Emergent Force}

\begin{figure}
  \centering
  \includegraphics[height=.3\textwidth]{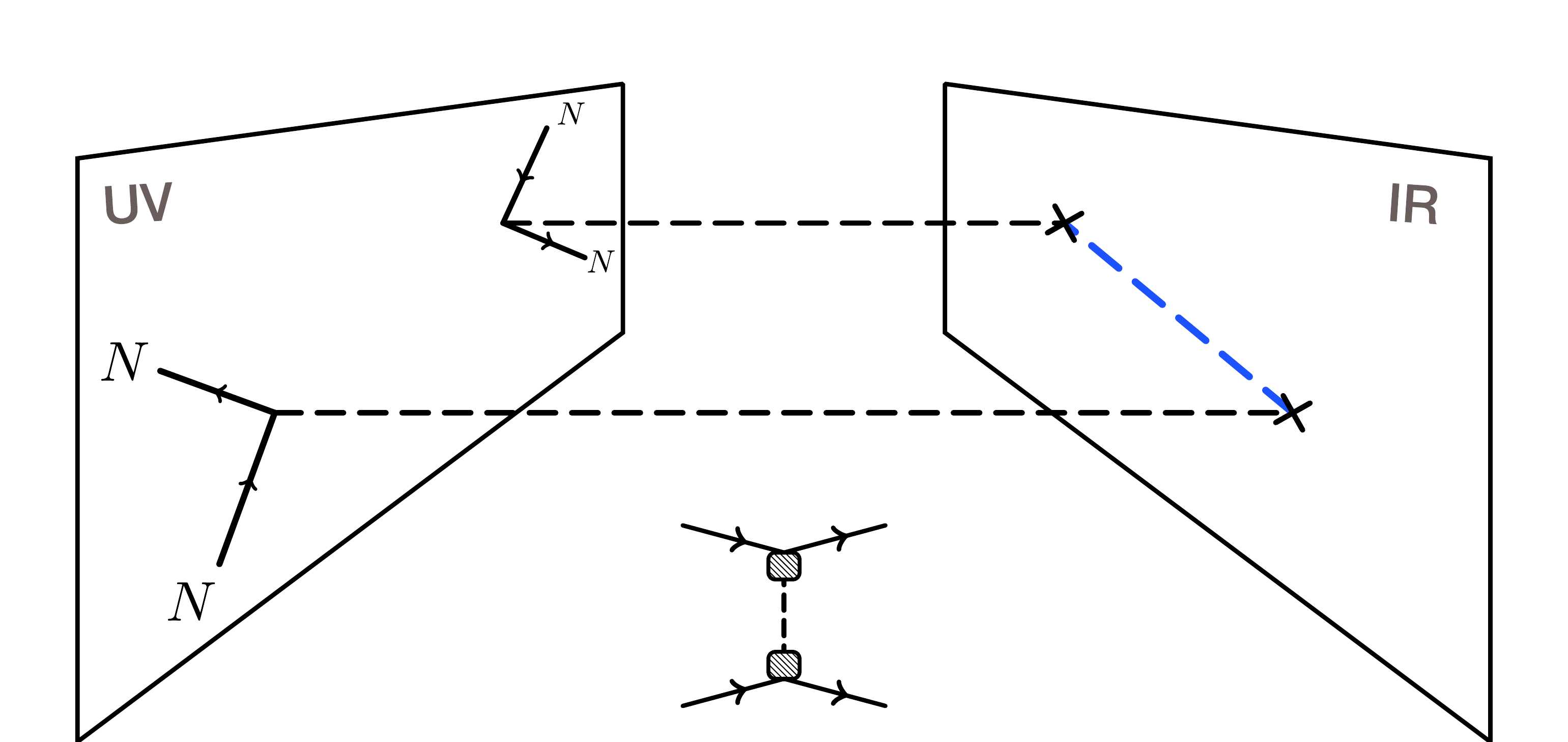}
  \caption{Diagram in 5D anti-de Sitter space giving rise to the emergent force. Blue line shows the propagator of the IR brane localized field. Crosses indicate mass mixing.}
  \label{fig:emergent}
\end{figure}

This emergent force scenario can be realized more precisely in a five-dimensional model, as suggested by the AdS/CFT corresponddence. We sketch this in Fig.~\ref{fig:emergent}.

Consider the limit where the dark sector is a strongly-coupled gauge theory with large 't Hooft coupling and a large number of colors.
The deconfined picture is similar to the picture described in Section~\ref{sec:warped:conformal}.
The key difference is that, at  four-momentum of order of the confinement scale, an observer on the \acro{UV} brane should see the appearance  of a mass gap.
The {\acro{AdS}}/\acro{CFT} correspondence relates confinement to presence of an \acro{IR} brane located at $z_1=1/\mu>z_0$. 
This \acro{IR} brane encodes the spontaneous breaking of conformal symmetry in the infrared. States localized on the \acro{IR} brane are identified with states that emerge below the confinement scale. 

For example, a \acro{UV} observer probing the bulk at low four-momentum observes a standard 4D theory. As four-momentum is increased beyond $\mu$, the effect of the IR brane vanishes from the correlation functions because the bulk propagators are exponentially suppressed when probing the IR region $z>1/p$ \cite{Fichet:2019hkg}. Hence for $p\gg \mu$, the \acro{IR} region of the bulk becomes opaque to observers on the \acro{UV} brane. In that limit, the observer sees only an infinite AdS bulk,  as described in the conformal/warped force scenario of Section~\ref{sec:warped:conformal}.

This mechanism can realize many versions of an emergent force. For simplicity, we focus on an extension of the model in Section~\ref{sec:warped:conformal}. Assume a warped 5D AdS space with curvature $k$. The Standard Model is localized on the \acro{UV} brane at $z_0 = 1/k$ and interacts with a bulk scalar field $\Phi$. We introduce an \acro{IR} brane at $z_1=1/\mu$. There may be a \acro{IR}-localized mass term. This term is left as a free parameter and  does not crucially affect the results.  
 Let us assume the existence an \acro{IR}-localized operator that interacts linearly with $\Phi$:
\begin{align}
  S=S_5+
\int_{\rm IR  \, brane} d^4X  \sqrt{|\gamma|}
\left( {\cal L}_{\rm IR} +{\cal O}_{\rm IR} \Phi    - 
 \frac{m_{\rm IR}^2}{2} \Phi^2
\right) \,,
\label{eq:emergent:IR:action}
\end{align}
where $S_5$ is the 5D action in \eqref{eq:warped:S5}.

The bulk Feynman propagator is readily expressed in terms of Bessel functions of the first and second kinds,
\begin{align}
\Delta_p(z,z') &=
i\frac{\pi (k z)^2(kz')^2}{2k }   & 
\frac{
\left[\tilde Y^{\rm UV}_{\alpha}J_{\alpha}\left(pz_<\right)
- \tilde J^{\rm UV}_{\alpha} Y_{\alpha}\left(pz_<
\right)\right]\left[
\tilde Y^{\rm IR}_{\alpha}J_{\alpha}\left(pz_>\right)
- \tilde J^{\rm IR}_{\alpha} Y_{\alpha}\left(pz_>
\right)
\right]}
{\tilde J^{\rm UV}_{\alpha} \tilde  Y^{\rm IR}_{\alpha}
- \tilde  Y^{\rm UV}_{\alpha} \tilde  J^{\rm IR}_{\alpha}}\,,
\label{eq:emergent:undressed:propagator}
\end{align}
where we define
the extra-dimensional positions
$z_{<}\equiv{\rm min}(z,z,')$, $z_{>}\equiv{\rm max}(z,z,')$, and the boundary functions
\begin{align}
  \tilde J^{\rm UV}_{\alpha} 
  & \equiv
  \frac{p}{k} \, J_{\alpha-1}\left(\frac{p}{k}\right) 
  &
  \tilde J^{\rm IR}_{\alpha} 
  & \equiv 
  \frac{p}{\mu} \, J_{\alpha-1}\left(\frac{p}{\mu}\right) + b_{\rm IR} \,J_\alpha\left(\frac{p}{\mu}\right)\, ,
\end{align}
and similarly for the $\tilde{Y}_\alpha$ functions.

We assume that a 4D field $\varphi$ is present on the \acro{IR} brane and mixes with $\Phi$ such that
\begin{align}
  {\cal O}_{\rm IR}
  = 
 \omega \varphi \Phi 
  \, .
\end{align}
with brane mass mixing parameter 
\be
\omega \equiv  c_\varphi\frac{\mu^2}{\sqrt{k}} \,.
\ee
The magnitude of $\omega$ is set by dimensional analysis, such that  $c_\varphi \sim \mathcal O(1)$. 
Since this is a bilinear interaction, the presence of the \acro{IR} brane field $\varphi$ can be rigorously included in the bulk propagator using
the five-dimensional version of dressing, \textit{i.e.} a geometric series representation of the brane--bulk mass mixing.  The \acro{UV}-to-\acro{UV} brane propagator dressed by \acro{IR} brane insertions is 
\begin{align}
  \Delta^{\rm dr}_p(z_0,z_0)
  &= 
  \Delta^{\rm}_p(z_0,z_0)
  - 
  \left[\Delta^{\rm }_p(z_0,z_1)\right]^2
  \frac{i\,\omega^2}{ p^2-m^2_{\varphi} } 
  \,. 
  \label{eq:Delta_dr}
\end{align}
We have introduced the $\varphi$  mass as
\begin{align}
 m^2_{\varphi}
 &=
 m^2_{\varphi,0} - i\,\omega^2 \Delta^{\rm }(m_{\varphi,0};z_1,z_1) \, ,
\end{align}
where $m^2_{\varphi,0}$ is a bare mass term and the $\Delta^{\rm }(p;z_1,z_1)$ term encodes the contribution to the mass from Kaluza--Klein modes. From dimensional analysis, both terms have a typical magnitude $\mathcal O(\mu^2)$. 
The first term of \eqref{eq:Delta_dr} corresponds to the undressed propagator. The second term encodes the effect of the \acro{IR} brane and  generates the emergent force in the infrared.

\subsection{Potential}

The spatial potential follows from the dressed propagator, \eqref{eq:Delta_dr}. However, the Fourier transform of this expression is only analytically tractable in the high- and low-energy limits. 
Let us consider the limits with space-like four-momentum $p=i |{\bf q}|\equiv i q$, as needed for the $t$-channel diagram that generates the potential. 
For $q\gg \mu$,
\begin{align}
\Delta_p(z_0,z_0)
  &\approx -i
     \frac{\Gamma(\alpha)}{\Gamma(1-\alpha)}\frac{2k}{q^2}\left(\frac{q}{2k}\right)^{2-2\alpha}
     \label{eq:Delta_undr}
     \\
\Delta_p(z_0,z_1)
  &\approx  -i \frac{\sqrt{2\pi}}{\Gamma(1-\alpha)}\frac{2k}{q^{3/2}\sqrt{\mu}} \left(\frac{q}{2k}\right)^{1-\alpha} e^{-q/\mu} 
  \\
  \Delta_p(z_1,z_1)
  &\approx -i\frac{k}{q\mu}\tanh(q/\mu) \,.
\end{align}
In the $q\ll \mu$ limit,
\begin{align}
  \Delta_p(z_0,z_0)
  &\approx - i  \frac{2\alpha+b_{\rm IR}}{2\alpha\, b_{\rm IR}}\frac{k}{\mu^2}\left(\frac{\mu}{k}\right)^{2-2\alpha}
  \\
  \Delta_p(z_0,z_1)
  &\approx - i \frac{k}{b_{\rm IR}\,\mu^2} \left(\frac{\mu}{k}\right)^{1-\alpha}
  \\
  \Delta_p(z_1,z_1)
  &\approx - i \frac{k}{b_{\rm IR}\,\mu^2}
  \,.
\end{align}

The behavior of the potential can be understood from these limits. 
   In the $q \ll \mu$ limit, all $\Delta$'s are constant and   only the \acro{IR}-brane scalar remains in the spectrum. The dressed propagator takes the form of a 4D massive propagator, thereby generating a standard 4D Yukawa force. 
  Equivalently, this is the behavior of the 4D effective theory whose Kaluza--Klein modes are integrated out.
   In the $q \gg \mu$ limit, the $\Delta^{\rm }(p;z_0,z_1)$ propagators are exponentially suppressed. Only the  $\Delta^{\rm }(p;z_0,z_0)$ term remains, reproducing the conformal scenario with \textit{no} \acro{IR} brane described in Section~\ref{sec:warped:conformal}.  
In the 4D CFT interpretation, this is because the composite states have typical size $1/ \mu$ and are not seen by probes with virtuality much higher than $\mu$. 

For the scalar nucleon operators, $\mathcal O^S$, the potential is:
\begin{align} \label{eq:VSem}
  V_S(r) &\approx
  \begin{dcases}
    -\frac{c^2_\varphi\lambda^2}{b_{\rm IR}^2}
    \left(\frac{\mu}{k}\right)^{2-2\alpha} 
    \frac{e^{-m_\varphi r}}{4\pi r}
    &
    {\rm if}\quad r\gg 1/\mu
 \\
    - \frac{\lambda^2 }{2\pi^{3/2}} 
    \frac{\Gamma(3/2-\alpha)}{\Gamma(1-\alpha)} 
    \frac{1}{r} 
    \frac{1}{(kr)^{2-2\alpha}}\,
  &
  {\rm if}\quad r\ll 1/\mu
\end{dcases} 
\end{align}
Analogously, for the pseudoscalar nucleon operators, $\mathcal O^P$, the potential is:
\begin{align} \label{eq:VPem}
  V_P({\bm r})
  &\approx
  \begin{dcases}
 \frac{c^2_\varphi\lambda^2}{b_{\rm IR}^2}
 \left(\frac{\mu}{k}\right)^{2-2\alpha}
 \frac{(\vec{\sigma_1}\cdot \vec{\nabla})(\vec{\sigma_2}\cdot \vec{\nabla})}{4m_N^2}
 \frac{e^{-m_\varphi r}}{4\pi r}
 & {\rm if}\quad r\gg 1/\mu
 \\
  \frac{\lambda^2 }{2\pi^{3/2}} 
  \frac{\Gamma(3/2-\alpha)}{\Gamma(1-\alpha)} 
  \frac{(\vec{\sigma_1}\cdot \vec{\nabla})(\vec{\sigma_2}\cdot \vec{\nabla})}{4m_N^2}
  \frac{1}{r} 
  \frac{1}{(kr)^{2-2\alpha}}\,
  & {\rm if}\quad r\ll 1/\mu
\end{dcases}
\ .
\end{align}

Just like the case of an emergent force from a confining non-Abelian sector in Section ~\ref{se:nonab}, there  is again an approximate continuity between the long and short distance regimes. This can be seen in  the spatial potentials (\ref{eq:VSem}--\ref{eq:VPem}), and also at the level in the propagator itself by comparing the undressed and emergent parts of \eqref{eq:Delta_dr} at momentum of order $\mu$.

Despite this approximate continuity, there may be a smooth kink around $r\sim 1/\mu$ if the emergent force is enhanced. This can happen for sufficiently large $\omega$. In our simple scalar model this requires some tuning to keep  $m_\varphi$ small. 
The emergent force can also be enhanced if $\varphi$ has some multiplicity, $n_{\varphi}$, in which case the emergent component of the force is increased by $n_{\varphi}$. Note however that $n_{\varphi}$ cannot be arbitrarily increased
because the  local strong coupling scale decreases as $n_{\varphi}^{-1/2}$.

Finally, it is instructive to estimate how the emergent force vanishes at short distances. 
In the  $q\gg \mu$ regime, the emergent contribution is
\begin{align}
\Delta^{\rm dr}_p(z_0,z_0) 
&\supset \frac{k\mu^3}{q^5} \left(\frac{q}{k}  \right)^{2-2\alpha}e^{-2q/\mu}\,,
\end{align}
where we have dropped all the $\mathcal O(1)$ factors for clarity.
This approximation is valid for momenta larger than $\mu$. We thus take 
$q>\tilde \mu$ where $\tilde \mu$ is an IR cutoff that may be taken to be around $\mathcal O(\text{few})\times \mu$. The Fourier transform integrates over a momentum range 
$q\in [\tilde \mu,\infty]$.
Expanding in small $r\tilde \mu$ shows that the emergent component of the potential behaves as
\begin{align}
\left.V(r) \right|_{r\ll 1/\mu}
\sim  c^2_\varphi \lambda^2  e^{-2 \tilde \mu / \mu}  \, \frac{\mu^4 r^2}{k}
\left(\frac{{\tilde \mu}}{k}\right)^{^{1-2\alpha}} + ~ \cdots \ .
 \end{align}
This limit explicitly shows how the emergent force vanishes at short distance. This is only a crude approximation as the overall magnitude strongly depends on $\tilde \mu$. 
An exact, numerical calculation of the holographic emergent force is beyond the scope of the present study.

\section{Conclusion}

Hidden (dark) sectors with new particles may generate long-range forces between visible sector matter.
This manuscript examines exotic long-range forces that differ from  the Yukawa-like forces generated from single-particle exchange. We present three classes of exotic forces.  

{Quantum forces}  come from the loop-level exchange of pairs of dark sector particles. 
They are described by an effective theory and may themselves be the dark matter.
We present the spin-dependent potentials including the spin- and orientation-averaged limits. As an example, we show the constraints on a light dark sector imposed from NMR bounds on the anomalous J-coupling of deuterium.

{Conformal forces} arise when visible particles couple to a  dark sector  with conformal symmetry. 
Such forces are also generated in the case of a ``warped dark sector,'' which by the AdS/CFT correspondence is a five-dimensional realization of the conformal dark sector. These forces have non-integer behavior in~$r$. 

{Emergent forces} are induced by effective degrees of freedom arising in the infrared. We presented a qualitative picture in a 4D strongly-interacting dark sector, and a quantitative result from a specific realization of this scenario in a slice of AdS$_5$. In the AdS model, the emergent force comes from 
an IR brane-localized degree of freedom that becomes invisible to the UV-localized nucleons at short distances. 

As an aside we classify the behavior of spin-dependent and spin-independent forces, for Yukawa and exotic cases, and for the ordinary, spin-averaged, and orientation-averaged cases. Such an analysis is required to form a coherent vision of existing and upcoming experimental prospects. 
We point out that in the orientation-averaged limit, the  Yukawa forces are suppressed as a result of Gauss' law. A similar effect also occurs for the tensor force upon spin-averaging. This behavior is not true for exotic forces. Yukawa forces are thus non-generic compared to exotic forces.   

It follows that experiments that use disordered phases of matter are particularly 
appropriate for searching for exotic spin-dependent forces. NMR-based experiments are one such type of setup. 
We find that searches for spin/velocity-dependent forces are especially sensitive to exotic tensor potentials.

 \section*{Acknowledgments}

We thank Michael Snow and Christopher Haddock for providing information about the Neutron Spin Rotation experiment. We also thank Brian Shuve and the Harvey Mudd College Physics Department for providing space for our group to collaborate. We thank the 2019 SoCal BSM workshop at UC Irvine for its hospitality while this manuscript was completed between sessions.

A.C.~is supported by the National Science Foundation Graduate Research Fellowship Program under Grant No.~1840991.
  S.F.~is supported by the S\~ao Paulo Research Foundation (FAPESP) under grants \#2011/11973, \#2014/21477-2 and \#2018/11721-4, and funded in part by the Gordon and Betty Moore Foundation through a Fundamental Physics Innovation Visitor Award (Grant GBMF6210).
P.T.~thanks the Aspen Center for Physics ({NSF} grant \#1066293) and the Kavli Institute for Theoretical Physics ({NSF} grant {PHY}-1748958) for their hospitality while this work was completed.
%

\appendix

\section{The Dipole Potential}
\label{app:dipole}

The four operators ${\cal O}_N^{\rm S,P,A,V}$ can couple to a dark mediator with renormalizable couplings.
In contrast, the tensor operator ${\cal O}_N^{\rm T}$ can only couple to other fields through a higher-dimensional operator. This is a consequence of its two Lorentz indexes. Since $\sigma_{\mu\nu}$ is antisymmetric, the only operator available is a field strength tensor $X_{\mu\nu}$ which couples to the tensor operator as 
\begin{align}
 \frac{m_N}{\Lambda^2} \bar N \sigma_{\mu\nu} N X^{\mu\nu}
&&
 \frac{m_N}{\Lambda^2} \bar N \sigma_{\mu\nu} N \tilde X^{\mu\nu}\label{eq:darkdipole}\,.
\end{align}
It is natural to assume that $X_{\mu \nu}$ is the field strength of a hidden gauge group.  The operators in \eqref{eq:darkdipole} then describe \textit{dark} magnetic and electric dipole moments. We assume the gauge group is Abelian and denote it $U(1)_X$. We refer to the gauge boson as the dark photon.

The Standard Model particles themselves may have hidden charge. 
This possibility is highly constrained due to the chiral structure of the Standard Model: either new chiral fermions must be carefully introduced to cancel anomalies, or Wess--Zumino terms are generated in the low-energy theory. 
A less constrained possibility is that all Standard Model fields are singlets under $U(1)_X$. In that case, the dark photon may have a kinetic mixing with the visible photon~\cite{Holdom:1986eq,Holdom:1985ag}. This mixing is typically loop induced so that the dark photon has a small coupling to visible electric currents.

Even without kinetic mixing,  visible sector fields can interact with the dark photon through multipole operators. Complex and Dirac fields can have dipoles, and self-conjugate fields can have dark polarizability.~\footnote{See \textit{e.g.} \cite{Fichet:2016clq} for more details on polarizability operators in the context of a dark sector. } 
In particular, nucleons (or quarks), have dark dipole operators like those in \eqref{eq:darkdipole}.  
Two ways to generate a dark polarization are: (i) The polarization may be induced  by  loops of heavy particles coupled to the Standard Model.  (ii) The polarization could be a consequence of the compositeness of Standard Model particles if the underlying constituents are charged under $U(1)_X$. In the latter case, the dipole moment is a low-energy manifestation of the internal structure of the Standard Model particle. This is analogous to the electromagnetic moments of hadrons.  
In this  ``dark dipole scenario,'' some amount of photon--dark photon kinetic mixing should also be present, at least as a result of loops contributing to $F^{\mu\nu}-X^{\mu\nu}$ mixing.    However, this loop-induced mixing via Standard Model fields can be expected to be small.

If the hidden gauge boson is sufficiently light, the nucleon dipole operator induces a spin-dependent force of tensor-type:
\begin{align}
  V_\text{T}({\bm r}) &= \frac{-4m_N^2 [(\vec{\sigma_1}\cdot\vec{\sigma_2})\nabla^2-(\vec{\sigma_1}\cdot\vec{\nabla})(\vec{\sigma_2}\cdot\vec{\nabla})]}{\Lambda^4}\left(\frac{e^{-mr}}{4\pi r}\right) \ .
\end{align}
Upon spin averaging $\vec{\sigma_2}$, the dominant piece is
\begin{align}
V_{\text{T}}({\bm r})&=\frac{\{ \mathbbm{1}_2[(\vec{p_1}+\vec{p_1}')\times \vec{\sigma_1}]\cdot\vec{\nabla}+ \mathbbm{1}_1 \mathbbm{1}_2\nabla^2\}}{\Lambda^4}\nabla^2\left(\frac{e^{-mr}}{4\pi r}\right) \ .
\end{align}
If, instead, the dark photon is heavy, then \eqref{eq:darkdipole} generates the $\bar N \sigma_{\mu\nu} N {\cal O}^{\mu\nu}_{\rm DS}$
tensor interaction. This operator  induces a quantum force,  presented in the analysis of Section~\ref{sec:quantum}. We leave further study of this ``dipole portal'' scenario for future work.

\section{Fourier Transforms and Effective Theory \label{app:FT_EFT}}

While the technique to derive a non-relativistic potential from a field theory amplitude is not new (see {e.g.}~\cite{Feinberg:1989ps}), some aspects related to the effective theory framework are usually left implicit and deserve clarification.~\footnote{This appendix is based on discussions between S.~F., G.~von~Gersdorff, and E.~Ponton.} 

The integral in the Fourier transform \eqref{eq:potential:from:amplitude}  
spans three-momenta up to infinity. However, whenever working within a low-energy effective theory, momenta higher than the effective theory cutoff $\Lambda$ should not be used in a calculation because it probes physics beyond the validity of the theory. For finite, low-energy predictions---like the potentials studied here---the details of the momentum truncation are ultraviolet details that should have negligible impact.

However, this leads to an apparent paradox. Amplitudes arising in the effective theory can grow with energy with a polynomial form such that the integrand in \eqref{eq:potential:from:amplitude}  takes typically the form $e^{iq r} q^{n} \ln q^2 $, with $n>0$, 
$q\equiv |\mathbf{q}|$, $r\equiv |\mathbf{r}|$. 
An integral up to infinity diverges and requires an ultraviolet cutoff.
One may impose the cutoff by introducing a step function $\Theta(|\mathbf{q}|<\Lambda)$ in the integrand.
In the presence of this factor the integral produces oscillating $\Lambda$-dependent terms such as $\Lambda^n \sin (\Lambda r)$ in addition to $\Lambda$-independent terms from the low-energy region of the Fourier integral. These $\Lambda$ terms are unsuppressed and do not vanish in the $\Lambda\rightarrow \infty$ limit. 

The same paradox occurs if one attempts an analytic continuation to transform the integral along the real line as an integral over the branch cut of the amplitude, as described in Appendix~\ref{app:pots} and used in Section~\ref{sec:quantum}. In that case,  the integral along the branch cut provides the universal long-distance contribution, while the integral over the large arcs of radius $\Lambda$ needed to close the contour gives rise to the $\Lambda$-dependent oscillating terms described above. 

These  $\Lambda$-dependent contributions originate from the fact that a hard cutoff factor $\Theta(|\mathbf{q}|<\Lambda)$ 
 introduces a non-analyticity at $|\mathbf{q}|=\Lambda$ because it is not continuous across this boundary. The $\Lambda$-dependent contributions are thus  artifacts of the truncation of momentum space. 
The solution to the paradox is then clear: A smooth cutoff should be used in order to avoid the spurious $\Lambda$-dependent contributions. 

Such smooth cutoff is conveniently implemented by convolving the step function with a smooth distribution, $\pi$. For example:
\begin{align}
 \Theta(|\mathbf{q}|<\Lambda)
&\to 
\int d\xi\, \Theta(|\mathbf{q}|<\Lambda+\xi)\pi(\xi)
&
\pi(\xi)&=\frac{1}{\sqrt{2\pi}\sigma}e^{-\xi^2/(2 \sigma^2)} \, ,
\end{align}
where  $\sigma\ll \Lambda$ is the width of the smoothing function.   
 The Fourier integral takes the form
\be
\int \frac{d^3 q}{(2\pi)^3} e^{i\vec{q}\cdot \vec{r}}  \mathcal{M}_{IJ} \int d\xi\, \Theta(|\mathbf{q}|<\Lambda+\xi)\pi(\xi) \,.
\ee
The $\xi$ integral is most conveniently performed after the Fourier transform.
The $\Lambda$-dependent contribution to the potential is  exponentially suppressed by a factor $e^{-r^2\sigma^2}$, thereby leaving the universal long-distance contribution as the main contribution to the potential.

\section{Calculation of the Quantum Potentials}
\label{app:pots}

We present additional details for the calculation of the quantum potentials in Section \ref{sec:quantum}.

\subsection{Loop Calculation}

The relevant one-loop amplitudes for the operators of Section \ref{sec:quantum} are:
\begin{align}
    i\mathcal{M}_a^0
    &=
    \frac{\eta}{\Lambda^4}(\bar{u}_{p_1'}\gamma^\mu \gamma^5 u_{p_1}\bar{u}_{p_2'}\gamma^\nu \gamma^5 u_{p_2})\int \frac{d^4k}{(2\pi)^4} \frac{2k_\mu+q_\mu}{k^2 - m^2 }\frac{2k_\nu+q_\nu}{(q + k)^2 - m^2 }
\\
    i\mathcal{M}_b^0
    &=
    \frac{2^{\eta-1}}{\Lambda^2}(\bar{u}_{p_1'}i\gamma^5 u_{p_1}\bar{u}_{p_2'}i\gamma^5 u_{p_2})\int \frac{d^4k}{(2\pi)^4}\frac{1}{k^2 - m^2 }\frac{1}{(q + k)^2 - m^2 }
\\
    i\mathcal{M}_c^0
    &=
    \frac{2^{\eta-1}}{\Lambda^6}(\bar{u}_{p_1'}i\gamma^5 u_{p_1}\bar{u}_{p_2'}i\gamma^5 u_{p_2})\int \frac{d^4k}{(2\pi)^4}\frac{k^2 + k\cdot q}{k^2 - m^2 }\frac{k^2 + k\cdot q}{(q + k)^2 - m^2 }
\\
    i\mathcal{M}_d^0
    &=
    \frac{\eta}{\Lambda^6}(\bar{u}_{p_1'}\sigma^{\mu\nu} u_{p_1}\bar{u}_{p_2'}\sigma^{\alpha\beta} u_{p_2} )\int \frac{d^4k}{(2\pi)^4} \frac{k_\mu(k+q)_\nu}{k^2 - m^2 }\frac{k_\alpha (k+q)_\beta}{(q+k)^2 - m^2 } \\
    i\mathcal{M}_a^\frac{1}{2}
    &=
    \frac{-2^{\eta-1}}{\Lambda^4}(\bar{u}_{p_1'}i\gamma^5 u_{p_1}\bar{u}_{p_2'}i\gamma^5 u_{p_2})\int \frac{d^4k}{(2\pi)^4} \mathrm{Tr}\left[\frac{(\slashed{k}+m)i\gamma^5}{k^2 - m^2 }\frac{(\slashed{q} + \slashed{k}+m)i\gamma^5}{(q + k)^2 - m^2 }\right]
\\
    i\mathcal{M}_b^\frac{1}{2}
    &=
    \frac{-2^{\eta-1}}{\Lambda^4}(\bar{u}_{p_1'}i\gamma^5 u_{p_1}\bar{u}_{p_2'}i\gamma^5 u_{p_2})\int \frac{d^4k}{(2\pi)^4} \mathrm{Tr} \left[\frac{(\slashed{k}+m)}{k^2 - m^2 }\frac{(\slashed{q} + \slashed{k}+m)}{(q + k)^2 - m^2 }\right]
\\
    i\mathcal{M}_c^\frac{1}{2}
    &=
    \frac{-2^{\eta-1}}{\Lambda^4}(\bar{u}_{p_1'}\gamma^\mu\gamma^5 u_{p_1}\bar{u}_{p_2'}\gamma^\nu\gamma^5 u_{p_2})\int \frac{d^4k}{(2\pi)^4} \mathrm{Tr}\left[\frac{(\slashed{k}+m)\gamma_\mu\gamma^5}{k^2 - m^2}\frac{(\slashed{q} + \slashed{k}+m)\gamma_\nu\gamma^5}{(q + k)^2 - m^2 }\right]
\\
    i\mathcal{M}_d^\frac{1}{2}
    &=
    \frac{-\eta}{\Lambda^4}(\bar{u}_{p_1'}\gamma^\mu\gamma^5 u_{p_1}\bar{u}_{p_2'}\gamma^\nu\gamma^5 u_{p_2})\int \frac{d^4k}{(2\pi)^4} \mathrm{Tr}\left[\frac{(\slashed{k}+m)\gamma_\mu}{k^2 - m^2 }\frac{(\slashed{q} + \slashed{k}+m)\gamma_\nu}{(q + k)^2 - m^2 }\right]
\\
  i\mathcal{M}_e^\frac{1}{2} 
  &=
  \frac{-\eta}{\Lambda^4}(\bar{u}_{p_1'}\sigma^{\mu\nu} u_{p_1}\bar{u}_{p_2'}\sigma^{\alpha\beta} u_{p_2} )\int \frac{d^4k}{(2\pi)^4}\mathrm{Tr}\left[\frac{(\slashed{k}+m)\sigma_{\mu\nu}}{k^2 - m^2 } \frac{(\slashed{q} + \slashed{k}+m)\sigma_{\alpha\beta}}{(q + k)^2 - m^2}\right]
  \end{align}\begin{align}
  i\mathcal{M}_a^1 
  &=
  \frac{2^{\eta+3}}{\Lambda^6}(\bar{u}_{p_1'}i\gamma^5 u_{p_1}\bar{u}_{p_2'}i\gamma^5 u_{p_2})\int \frac{d^4k}{(2\pi)^4}\frac{1}{k^2 - m^2 }\frac{2(k\cdot(k+q))^2+k^2(k+q)^2}{(q + k)^2 - m^2}
\\  
  i\mathcal{M}_b^1 
  &=
  \frac{2^{\eta+4}}{\Lambda^6}(\bar{u}_{p_1'}i\gamma^5 u_{p_1}\bar{u}_{p_2'}i\gamma^5 u_{p_2})\int \frac{d^4k}{(2\pi)^4}\frac{1}{k^2 - m^2 }\frac{(k\cdot(k+q))^2-k^2(k+q)^2}{(q + k)^2 - m^2 }
  \ ,
\end{align}
where $q=p_1-p_1'=p_2'-p_2$. Unprimed momenta represent the initial states, and primed momenta represent the final states. We introduce Feynman parameters to simplify the integral in the usual way. The resulting  integrals are
\begin{align}
    \int \frac{d^4l}{(2\pi)^4}\frac{1}{(l^2 - \Delta)^2} 
    &\longrightarrow
    \frac{-i}{(4\pi)^2}\ln \left(\frac{\Delta}{\Lambda^2}\right)
\\
  \int \frac{d^4l}{(2\pi)^4}\frac{l^2}{(l^2 - \Delta)^2} 
  &\longrightarrow
  \frac{-2i\Delta}{(4\pi)^2}\ln \left(\frac{\Delta}{\Lambda^2}\right)
\\  
  \int \frac{d^4l}{(2\pi)^4}\frac{(l^2)^2}{(l^2 - \Delta)^2} 
  &\longrightarrow
  \frac{-3i\Delta^2}{(4\pi)^2}\ln \left(\frac{\Delta}{\Lambda^2}\right)
   \end{align}
   with $\Delta=m^2-x(1-x)q^2$. 
The amplitudes can then be written in a basis of integrals over the Feynman parameters,
\begin{align}
    f_n &= \int_{0}^{1}dx (x(1-x))^n\ln \left(\frac{\Delta}{\Lambda^2}\right) \ .
\end{align}
$\ln(y)$ has a branch cut along the real axis for $y<0$. The discontinuity in $f_n$ due to this branch cut is given by
\begin{align}
	\mathrm{Disc}(f_n) &= 2\pi i\int_{x_-}^{x_+}dx (x(1-x))^n
  &
	x_\pm &= \frac{1}{2}\pm\frac{1}{2q}\sqrt{q^2 -4m^2}
  \ .
\end{align}

\subsection{Amplitude to Spatial Potential}

The spatial potential is a Fourier transform of the relativistic scattering amplitude $\mathcal M$,
\begin{align}
  V({\bm r}) 
  &= 
  \int\frac{d^3q}{(2\pi)^3} 
  \frac{-\mathcal{M}(\vec{q})}{4m_N^2}
  e^{i\vec{q}\cdot\vec{r}}
  &
  \mathcal{M}(\vec{q}) 
  &=
  \sum_A
  \mathcal S_A(\vec{q})
  f_A(|\vec{q}|)
  \ .
\end{align}
Here $A$ indexes possible tensor structures in spin space as carried by the factors $\mathcal S_A(\vec{q})$. Observe that $\mathcal S_A(\vec{q})$ may depend on $\vec{q}$ through $(\vec{q}\cdot\sigma_1)(\vec{q}\cdot\sigma_2)$. This is the only piece of the amplitude that may depend on $\vec{q}$ as a spatial vector rather than just its magnitude. Inside the Fourier transform, we may identify the transfer momentum with a gradient, $\vec{q} = -i\nabla$. This gives an expression for the potential that is a Fourier transform of a function that only depends on the magnitude, $\rho = |\vec{q}|$:
\begin{align}
  V({\bm r}) 
  &= 
  \sum_A
  \frac{-i
  \mathcal S_A(-i\nabla)
  }{4m_N^2}
  \int_{-\infty}^\infty 
  \frac{d\rho}{(2\pi)^2} 
  \rho f_A(\rho)
  e^{i\rho r} \ ,
\end{align}
where we have performed the angular integrals and have extended the radial integral to the entire real line. The remaining integral may be performed by analytic continuation into the complex plane, minding the branch cuts in the $f_A(\rho)$ functions along the imaginary $\rho$-axis starting at $\rho = 2im$. Deforming the integration contour then maps the integral to the discontinuity across this branch cut: 
\begin{align}
  \int_{-\infty}^\infty 
  \frac{d\rho}{(2\pi)^2} 
  \rho f_A(\rho)
  \frac{e^{i\rho r}}{r} 
  &=
  \int_{2im+\varepsilon}^{i\infty+\varepsilon}
  \frac{d\rho}{(2\pi)^2} 
  \rho f_A(\rho)
  \frac{e^{i\rho r}}{r}
  -
  \int_{2im-\varepsilon}^{i\infty-\varepsilon}
  \frac{d\rho}{(2\pi)^2} 
  \rho f_A(\rho)
  \frac{e^{i\rho r}}{r}
  \ .
\end{align}
Changing integration variables then yields:
\begin{align}
  V({\bm r}) 
  &= 
  \sum_A
  \frac{i
  \mathcal S_A(-i\nabla)
  }{4m_N^2}
  \int_{2m}^{\infty}
  \frac{d\lambda}{(2\pi)^2}\lambda
  \,
  \text{Disc}\left[f_A(\lambda)\right]
  \frac{e^{-\lambda r}}{r}
   \ .
\end{align}

To complete the remaining integral, we use
\begin{align}
  \int_{2m}^{\infty}d\lambda\sqrt{\lambda^2-4m^2}e^{-\lambda r}
  & =
  \frac{2m}{r}K_1(2mr)
\\
  \int_{2m}^{\infty}d\lambda\lambda^2\sqrt{\lambda^2-4m^2}e^{-\lambda r}
  & =
  \frac{8m^3}{r}K_1(2mr)+\frac{12m^2}{r^2}K_2(2mr)
\\
  \int_{2m}^{\infty}d\lambda\lambda^4\sqrt{\lambda^2-4m^2}e^{-\lambda r}
  & =
  \frac{32m^4}{r^2}K_2(2mr)+\left(\frac{120m^3}{r^3}+\frac{32m^5}{r}\right)K_3(2mr)
\\
  \int_{2m}^{\infty}d\lambda\lambda^6\sqrt{\lambda^2-4m^2}e^{-\lambda r}
  & =
  2m^8\left[\frac{K_1(2mr)}{2mr}+\frac{9K_2(2mr)}{(2mr)^2}+\frac{45K_3(2mr)}{(2mr)^3}+\frac{105K_4(2mr)}{(2mr)^4}\right]
\\
    \int_{2m}^{\infty}\frac{d\lambda}{\lambda^2}\sqrt{\lambda^2-4m^2}e^{-\lambda r}
    & =
     \frac{1}{4m^2r^2}\left(4+\pi m^3r^3+\pi^2m^2r^2G(m^2r^2)\right) \ .
\end{align}
$G(m^2r^2)$ is shorthand for one of the Meijer G--functions,
\begin{equation}
	G(m^2r^2) \equiv G_{2,4}^{2,0}\left(m^2 r^2\bigg|
	{\begin{array}{c}
	\frac{1}{2},\frac{3}{2} \\
	0,0,\frac{1}{2},\frac{1}{2} \\
	\end{array}}
	\right) \ .
\end{equation}
The orientation-averaged form for the potentials is equivalent to the replacement
\begin{equation}
	\partial_i \partial_j \longrightarrow \frac{1}{3}\delta_{ij}\nabla^2  \ .
\end{equation}

\section{Non-Relativistic Spinor Limits}
\label{app:NRlims}

For convenience, we present results of spinor contractions to leading order in the non-relativistic limit. Latin indices refer to spatial directions.

\subsection{Both Sources Polarized}
\begin{align}
  \bar{u}_{p_1'}u_{p_1} \bar{u}_{p_2'} u_{p_2} 
  &\approx
  4m_N^2\mathbbm{1}_1 \mathbbm{1}_2
\\
  \bar{u}_{p_1'}\gamma^\mu u_{p_1} \bar{u}_{p_2'}\gamma^\nu u_{p_2} 
  &\approx
  4m_N^2\delta_0^\mu\delta_0^\nu\mathbbm{1}_1 \mathbbm{1}_2
\\    
  \bar{u}_{p_1'}\sigma^{\mu\nu} u_{p_1} \bar{u}_{p_2'}\sigma^{\rho\lambda} u_{p_2} 
  &\approx
  4m_N^2\varepsilon^{ijk}\varepsilon^{lmn}\delta_i^\mu\delta_j^\nu\delta_l^\rho\delta_m^\lambda\sigma_1^k\sigma_2^n
\\    
  \bar{u}_{p_1'}\gamma^\mu\gamma^5 u_{p_1} \bar{u}_{p_2'}\gamma^\nu\gamma^5 u_{p_2} 
  &\approx
  4m_N^2\delta_i^\mu\delta_j^\nu\sigma_1^i\sigma_2^j
\\    
  \bar{u}_{p_1'}i\gamma^5 u_{p_1} \bar{u}_{p_2'}i\gamma^5 u_{p_2} 
  &\approx
  q_j q_k \sigma_1^j \sigma_2^k \ .
\end{align}

\subsection{One Source Polarized, Other Unpolarized}

The above results change when one source of nucleons is unpolarized. Take $\vec{\sigma_2}$ to represent the spin of the unpolarized nucleon current. The long-range potential is the average of the initial spins. The spin-independent $\mathcal O^S$ and $\mathcal O^V$ bilinears remain unchanged. The axial and pseudo-scalar combinations vanish at all order:
\begin{align}
	\bar{u}_{p_1'}\gamma^\mu\gamma^5 u_{p_1} \bar{u}_{p_2'}\gamma^\nu\gamma^5 u_{p_2} 
  &=0
  &
	\bar{u}_{p_1'}i\gamma^5 u_{p_1} \bar{u}_{p_2'}i\gamma^5 u_{p_2} 
  &=0 \ .
\end{align}
The tensor combination at leading order is
\begin{equation}
	\bar{u}_{p_1'}\sigma^{\mu\nu} u_{p_1} \bar{u}_{p_2'}\sigma^{\rho\lambda} u_{p_2} \approx  -iq_a \mathbbm{1}_2[((\vec{p_1}+\vec{p_1'})\times \vec{\sigma_1})_i-iq_i\mathbbm{1}_1](\delta^\mu_i\delta^\nu_0-\delta^\nu_i\delta^\mu_0)(\delta^\rho_a\delta^\lambda_0-\delta^\lambda_a\delta^\rho_0) \ .
\end{equation}

\bibliographystyle{utcaps} 	
\bibliography{biblio}

\end{document}